\documentclass[conference]{IEEEtran}

\ifCLASSINFOpdf
\else
\fi
\usepackage{url} 
\usepackage{graphicx}
\usepackage{booktabs}
\usepackage{amsmath}
\usepackage{subcaption} 
\usepackage{amssymb}
\usepackage{algorithm}
\usepackage{multirow}
\usepackage{algpseudocode}
\begin{document}
%
\title{Agent-Assisted Side-Channel Attacks on Non-Prefix KV Cache in RAG}


\author{
    \IEEEauthorblockN{He Sun}
    \IEEEauthorblockA{\textit{University of Science and Technology of China} \\
    hesun@mail.ustc.edu.cn}
    \and
    \IEEEauthorblockN{Shinan Liu}
    \IEEEauthorblockA{\textit{University of Hong Kong} \\
    shinan6@hku.hk}
    \and
    \IEEEauthorblockN{Siyuan Ma}
    \IEEEauthorblockA{\textit{Nanyang Technological University} \\
    MASI0004@e.ntu.edu.sg}
    \and
    \IEEEauthorblockN{Junhao Li}
    \IEEEauthorblockA{\textit{University of Science and Technology of China} \\
    lijh2125@mail.ustc.edu.cn}
    \and
    \IEEEauthorblockN{Mingjun Xiao}
    \IEEEauthorblockA{\textit{University of Science and Technology of China} \\
    xiaomj@ustc.edu.cn}
    \and
    \IEEEauthorblockN{Wenhao Jiang}
    \IEEEauthorblockA{\textit{Guangming Lab} \\
    cswhjiang@gmail.com}
}
	

%


\maketitle

\begin{abstract}
Modern Large Language Model (LLM) serving engines increasingly rely on Retrieval-Augmented Generation (RAG) and non-prefix Key-Value (KV) cache fusion to accelerate long-context, multi-tenant inference. While existing KV cache side-channel attacks require strict linear prefix alignment—rendering them ineffective against real-world RAG queries that contain unique, user-specific private prefixes—we uncover a critical class of structural vulnerabilities inherent to chunk-aware memory scheduling. We demonstrate that the deterministic micro-architectural mechanisms used to align and fuse disjoint memory chunks inadvertently leak a continuous \textit{Step-Wave} timing signature. Exploiting this physical observation, we introduce \textbf{SpliceLeak}, the first end-to-end side-channel attack targeting non-prefix KV cache fusion. SpliceLeak executes a systematic two-phase privacy breach: it first structurally fingerprints the exact length of hidden private prompts, and subsequently manipulates boundary collisions to extract exact semantic content token-by-token. Extensive evaluations on production-grade frameworks (vLLM integrated with LMCache) demonstrate that SpliceLeak achieves up to a 100\% extraction success rate in bounded-entropy scenarios. Driven by a deterministic $+104$ ms hardware latency void, the attack requires as few as 63 requests per token, piercing through realistic continuous batching noise. To resolve the inherent conflict between memory deduplication and security, we propose \textbf{SpliceDefense}, a bipartite mitigation framework consisting of Quantized Chunk Padding (QCP) and Constant-Time Boundary Fusion (CTBF). Our evaluations confirm that SpliceDefense effectively flattens the side-channel signal ($\Delta \text{TTFT} \approx 0$) with negligible throughput overhead, preserving the critical benefits of global cache sharing.
\end{abstract}

\section{Introduction} 
\label{sec:intro}

Large Language Models (LLMs) have fundamentally transformed natural language processing, demonstrating unprecedented capabilities across diverse tasks \cite{zhao2023survey,chang2024survey,oyelade2025survey}. However, deploying these models for long-context inference introduces severe computational and memory bottlenecks \cite{lee2024infinigen,sun2026hillinfer}. The prefill phase, characterized by massive quadratic attention matrix calculations, leads to unacceptably high Time-to-First-Token (TTFT) latency \cite{agrawal2024medha,sun2025breaking,wu2024loongserve,lin2024infinite}. Furthermore, when deployed as general-purpose assistants, LLMs frequently suffer from factual hallucinations due to a lack of domain-specific or up-to-date knowledge \cite{huang2025survey,liu2024survey}.

To counteract hallucinations, Retrieval-Augmented Generation (RAG) has emerged as the prevailing paradigm, incorporating external knowledge bases to ground responses in factual contexts \cite{zhao2024retrieval,ye2024r2ag,agarwal2025cache}. Concurrently, to alleviate the TTFT bottlenecks inherent in long-context processing, modern LLM serving engines extensively adopt Key-Value (KV) cache sharing \cite{kwon2023efficient,zheng2024sglang,zhong2024distserve,qin2025mooncake}. Foundational memory management techniques, such as PagedAttention \cite{kwon2023efficient} and RadixAttention \cite{zheng2024sglang}, enable the fine-grained retention and reuse of KV tensors. However, RAG workloads demand extreme flexibility due to their combinatorial nature. Consequently, state-of-the-art RAG serving frameworks (e.g., LMCache \cite{liu2025lmcache} or CacheBlend \cite{yao2025cacheblend}) have pioneered \textit{non-prefix KV cache fusion} mechanisms. By enabling the dynamic reuse and stitching of disjoint context chunks at arbitrary, non-prefix prompt positions, these advanced architectures drastically reduce the computational latency and memory footprint of complex, multi-document multi-tenant queries.

Naturally, the extensive sharing of physical memory states in multi-tenant cloud environments raises severe privacy concerns \cite{chu2025selective,zhang2026cryptogen,zeng2025mpcache}. Recent pioneering works, such as Shadow in the Cache \cite{luo2025shadow} and PROMPTPEEK \cite{wu2025know}, have demonstrated that attackers can exploit KV cache sharing as a timing side-channel to deduce prompt lengths and reconstruct sensitive user inputs. However, a fundamental architectural limitation of these existing methods is their strict reliance on \textit{linear prefix alignment} to establish a timing oracle. In realistic RAG scenarios, retrieved documents are invariably preceded by user-specific private prefixes, such as personalized system instructions or conversational histories. Even a single-token deviation at the prompt's root breaks this strict prefix alignment. Consequently, conventional caching mechanisms forcefully isolate the request, falling back to a full attention prefill. Because both correct and incorrect document guesses by the attacker incur this identical full-prefill latency, the macroscopic timing difference collapses entirely ($\Delta \text{TTFT} \approx 0$). This structural disappearance of the timing signal completely blinds traditional oracles, rendering existing prefix-based attacks fundamentally ineffective in RAG environments and creating a perilous false sense of security. 

In this paper, we demonstrate that this perceived security is fundamentally illusory. We reveal that the core micro-architectural optimizations driving RAG throughput—specifically, fixed-size chunk routing and non-prefix selective re-computation—inherently violate cross-tenant timing isolation. Because modern serving engines must strictly align memory blocks to stitch disjoint context chunks, they are forced to trigger a deterministic attention re-computation for any unaligned residual tokens at the fusion boundary.

We expose that this deterministic scheduling manifests as a high-fidelity micro-architectural side-channel. Through rigorous empirical profiling, we uncover the \textit{Step-Wave Effect}: rather than a discrete cache-miss penalty, the latency overhead strictly adheres to a bipartite linear model, governed by the fixed chunk routing cost (e.g., intercept) and the unaligned tail prefill cost (e.g., slope). By mathematically inverting this physical hardware behavior, an attacker completely circumvents the strict prefix isolation barrier, weaponizing the memory scheduler itself into a high-resolution, continuous length oracle.

Operationalizing this micro-architectural vulnerability, we present \textbf{SpliceLeak}, the first practical end-to-end side-channel exploit targeting non-prefix KV cache fusion in multi-tenant RAG systems. SpliceLeak orchestrates a highly deterministic two-phase privacy breach:
\begin{enumerate}
    \item \textbf{Phase I: Structural Fingerprinting.} By leveraging the Step-Wave Oracle, an attacker mathematically reconstructs the exact length of a victim's hidden prefix from a single TTFT measurement. This structural fingerprinting achieves single-token precision, drastically reducing the semantic entropy of the targeted prompt.
    \item \textbf{Phase II: Semantic Extraction.} Armed with this precise structural blueprint, the attacker deploys an \textit{LLM-driven Extraction Agent}. By dynamically generating high-probability vocabulary subsets and continuously monitoring the deterministic cache collision penalties at the fusion boundary, the agent iteratively reverse-engineers the victim's private semantics token-by-token in a continuous right-to-left traversal.
\end{enumerate}

Finally, recognizing the critical tension between aggressive memory deduplication and micro-architectural isolation, we propose \textbf{SpliceDefense}, a synergistic, two-stage mitigation framework. Moving beyond naive isolation paradigms that destroy cache sharing, SpliceDefense introduces Quantized Chunk Padding (QCP) to neutralize structural length oracles, and Constant-Time Boundary Fusion (CTBF) to eliminate semantic timing dips. Together, they effectively eradicate the side-channel Signal-to-Noise Ratio (SNR) while preserving the critical throughput benefits of global KV cache sharing.

In summary, our core contributions are:
\begin{itemize}
    \item \textbf{Vulnerability Discovery \& Theoretical Modeling:} We identify a fundamental micro-architectural vulnerability in modern non-prefix KV cache fusion mechanisms. We are the first to mathematically model the \textit{Step-Wave Effect} in chunk-aware memory scheduling, proving that current prefix-centric threat models severely underestimate the privacy risks in multi-tenant LLMaaS environments.
    
    \item \textbf{Novel Agentic Attack Framework (SpliceLeak):} We design and implement SpliceLeak, a robust two-phase attack that perfectly bypasses strict prefix isolation barriers. Empowered by an automated algorithmic extraction pipeline, we demonstrate its lethality across three distinct real-world scenarios: Confidential Input Extraction, Proprietary Template Reverse-Engineering, and Zero-Knowledge Privacy Reconstruction.
    
    \item \textbf{Comprehensive Empirical Evaluation:} We systematically evaluate SpliceLeak on production-grade serving architectures (vLLM with LMCache) utilizing advanced GQA/MHA mechanisms. Our bulk experiments reveal up to a 100\% extraction success rate across structured templates and bounded-search scenarios. Driven by a deterministic hardware latency void (e.g., $+104$ ms), SpliceLeak systematically extracts deeply private semantics utilizing as few as 63 requests per token, demonstrating high resilience against practical continuous batching noise.
    
    \item \textbf{Deployable Mitigations (SpliceDefense):} We resolve the inherent conflict between non-prefix memory deduplication and side-channel security by proposing SpliceDefense. Through real-world micro-architectural validation, we confirm that our coordinated QCP and CTBF mechanisms completely flatten the physical TTFT signal ($\Delta \text{TTFT} \approx 0$) with negligible throughput overhead, effectively restoring strict timing isolation and mitigating micro-architectural side-channels for RAG frameworks
\end{itemize}

\begin{figure*}[t]
    \centering
    \includegraphics[width=0.95\textwidth]{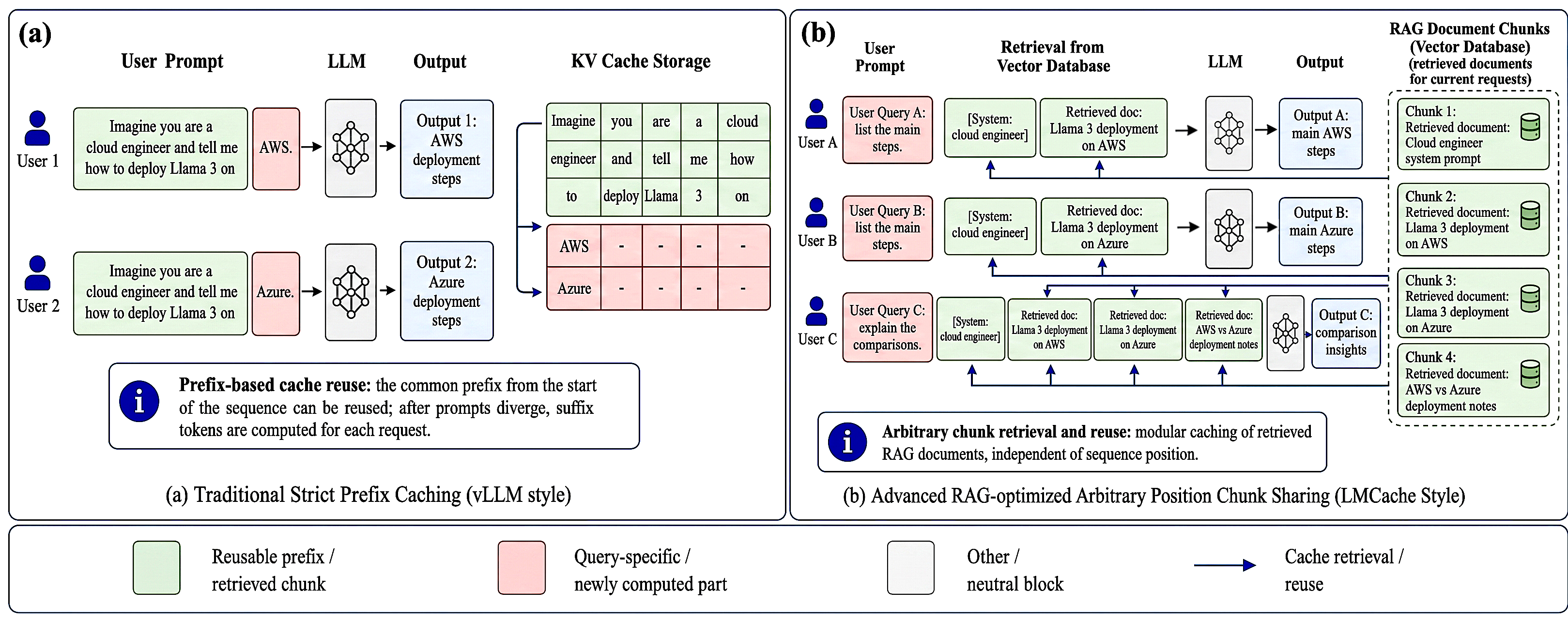}
    \caption{Comparison of KV cache sharing mechanisms in LLM serving and RAG-optimized serving. 
    (a) Traditional strict prefix caching, where the common prefix from the beginning of the sequence can be reused, while divergent suffix tokens are computed for each request. 
    (b) RAG-optimized arbitrary-position chunk sharing, where retrieved document chunks can be modularly reused independent of their positions in the prompt.}
    \label{fig:comparison_llm_rag}
    \vspace{-0.1in}
\end{figure*}
\section{Background}
\noindent\textbf{LLM Generation and Strict Prefix Caching.}
Modern LLM inference operates autoregressively, systematically divided into two distinct phases: prefill and decode \cite{zhao2023survey,chang2024survey}. During the prefill phase, the serving engine ingests the entire input sequence in parallel to compute the initial hidden states and materialize the Key-Value (KV) tensors. Because the computational complexity of the attention mechanism scales quadratically with the sequence length \cite{vaswani2017attention,beltagy2020longformer}, processing extensive contexts—such as those inherent to RAG workloads—requires massive, compute-bound matrix multiplications. Consequently, this prefill phase stands as the primary bottleneck dominating the Time-to-First-Token (TTFT) latency. To mitigate this severe computational overhead, state-of-the-art LLM serving frameworks, most notably vLLM \cite{kwon2023efficient} and SGLang \cite{zheng2024sglang}, have pioneered strict prefix caching mechanisms.

As illustrated in Figure \ref{fig:comparison_llm_rag}(a), when multiple requests share an identical, contiguous starting sequence (e.g., the shared instruction ``Imagine you are a cloud engineer...'' for both User 1 and User 2), the engine computes and retains the KV tensors for this common prefix (highlighted in green). When the prompts subsequently diverge (e.g., ``AWS'' versus ``Azure'', highlighted in red), the system only executes the computationally expensive prefill for the newly appended query-specific tokens. In the subsequent memory-bound decode phase, the model generates new tokens sequentially, utilizing the fully assembled KV cache to bypass redundant matrix multiplications.

\noindent\textbf{Continuous Batching and Scheduling Determinism.}
To maximize GPU utilization in multi-tenant environments, modern engines employ continuous batching (or iteration-level scheduling). Unlike static batching, continuous batching dynamically injects new requests into the active execution queue at the token-generation iteration boundary. Crucially, the scheduler inherently prioritizes requests whose KV states are already materialized in the cache. If a new request shares a cached prefix (as seen in Figure \ref{fig:comparison_llm_rag}(a)), it rapidly bypasses the extensive prefill stage and is prioritized to enter the decode phase. While this mechanism drastically optimizes overall system throughput, it inadvertently introduces a highly deterministic relative processing order among concurrent requests. This prioritization logic transforms a performance optimization into a measurable micro-architectural vulnerability, serving as the foundation for scheduling-based side-channel leakage \cite{wu2025know,luo2025shadow}.

\noindent\textbf{Retrieval-Augmented Generation (RAG) Serving Workflow.}
While LLMs demonstrate remarkable generative capabilities, they inherently suffer from hallucinations and knowledge cut-offs when responding to domain-specific or real-time queries \cite{liu2024survey,huang2025survey}. RAG architectures effectively mitigate these limitations by dynamically injecting external knowledge into the generation pipeline. As illustrated in Figure \ref{fig:comparison_llm_rag}(b), a standard RAG serving workflow comprises distinct retrieval and generation phases. When a user submits an initial query, the retriever searches a vector database to fetch the top-k most relevant document chunks.

Subsequently, the serving engine assembles a composite prompt for the LLM to generate the final output. Crucially, this composite prompt is highly structural and combinatorial. Rather than a simple linear text input, it is typically constructed by concatenating a foundational private system instruction (e.g., a designated persona like ``cloud engineer''), the retrieved modular chunks, and the specific user query. This combinatorial paradigm dictates that while the underlying retrieved documents are frequently shared and reused across multiple distinct user requests (e.g., Chunk 1 being utilized by both User A and User C), their absolute sequence positions and surrounding contextual tokens are highly dynamic. Consequently, modern RAG systems require advanced, position-independent chunk sharing mechanisms to efficiently cache and reuse these modular knowledge components \cite{yao2025cacheblend,liu2025lmcache}, fundamentally bypassing the restrictive linear requirements of traditional strict prefix caching.

\noindent\textbf{From Prefix Caching to Arbitrary Chunk Reuse.} 
To alleviate the severe TTFT bottlenecks inherent to the prefill phase, cloud providers implement KV cache sharing. Traditional frameworks (e.g., vLLM \cite{kwon2023efficient}, SGLang \cite{zheng2024sglang}) utilize strict prefix caching, which retains and shares the KV tensors of identical initial tokens across multiple requests. However, this architecture relies on a strict linear matching assumption: if a sequence diverges by even a single token—such as a personalized, private system instruction prepended to a user's prompt—the cache match instantly aborts. Consequently, standard prefix caching is fundamentally inadequate for multi-tenant RAG workloads, where highly reusable retrieved chunks are invariably preceded by distinct, user-specific private contexts.

To unlock the performance potential of RAG serving, next-generation frameworks (e.g., LMCache \cite{liu2025lmcache} or CacheBlend \cite{yao2025cacheblend}) have introduced \textit{non-prefix KV cache fusion}. This mechanism enables the arbitrary, position-independent reuse of retrieved document chunks. To manage GPU memory I/O efficiently, these systems partition cached documents into fixed-size KV blocks. When a disjoint, pre-cached chunk is fused into a new prompt at an arbitrary position, the system executes a specialized process known as KV Blending. Due to positional encoding shifts (e.g., RoPE) and cross-attention dependencies, the engine cannot simply concatenate memory pointers. Instead, it performs selective re-computation: it selectively routes and loads the maximally aligned full chunks directly from the shared memory pool, while forcing a full attention prefill for any unaligned residual tokens (i.e., the tail). 

While this selective re-computation drastically reduces latency compared to a full prefill, its computational overhead is strictly bipartite: a fixed memory routing cost per aligned chunk, and a linear computational penalty for the unaligned tail tokens modulo the chunk size. As we will demonstrate, this deterministic, chunk-aware scheduling logic inadvertently transforms the engine's prefill latency into a high-fidelity, micro-architectural timing fingerprint.
\begin{figure}[t]
    \centering
    \includegraphics[width=\linewidth]{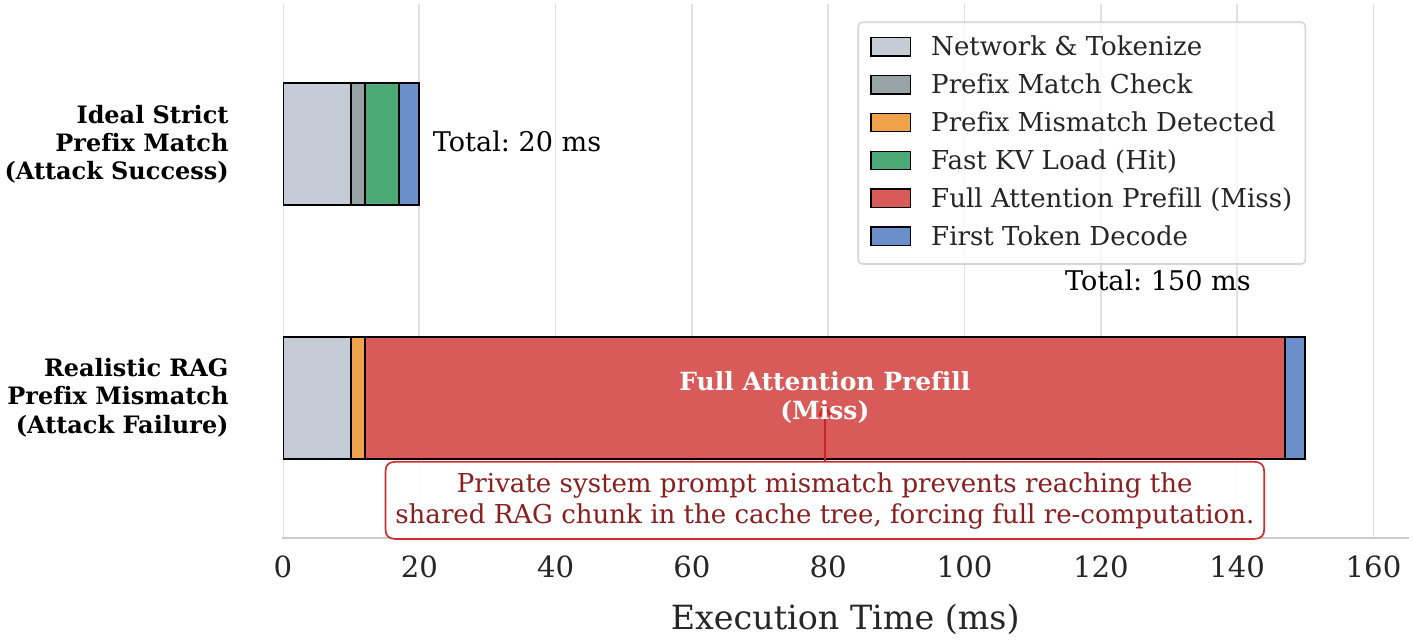}
    \caption{Micro-architectural execution timeline comparing an ideal prefix cache hit against a RAG prefix mismatch. The mismatch forces a fallback to the slow Full Attention Prefill, neutralizing the latency advantage of the cached chunk.}
    \label{fig:timeline_llm_rag}
\end{figure}

\section{Motivation: Limitations of Existing Works}
\label{sec:motivation}

\noindent\textbf{Mechanisms of Existing Side-Channel Attacks.} 
Recent pioneering studies, such as Shadow in the Cache \cite{luo2025shadow} and PROMPTPEEK \cite{wu2025know}, have successfully demonstrated that shared KV caches in multi-tenant LLM serving can be exploited as a timing side-channel. The core principle relies on treating the LLM serving engine as a Timing Oracle. By sending a carefully crafted probe prompt, the attacker measures the TTFT. A significantly low TTFT indicates a Cache Hit (i.e., the victim has queried the exact sequence), while a high TTFT indicates a Cache Miss. Attackers exploit this macroscopic timing difference ($\Delta T$) to reconstruct the victim's prompt token-by-token. However, a fundamental architectural prerequisite for this oracle to function is \textit{strict prefix alignment}. The attacker's probe must perfectly match the victim's prompt starting from the very first token to trigger the underlying PagedAttention/RadixTree hit.

\noindent\textbf{The Paradigm Shift in RAG Serving.} 
This strict prefix assumption structurally conflicts with the practical deployment of Retrieval-Augmented Generation. A standard RAG prompt is intrinsically composite: it begins with a user-specific private prefix $\mathcal{P}_{vic}$ (e.g., persona settings or conversational history), followed by the retrieved knowledge chunks $\mathcal{C}_{target}$, and concludes with the specific query. To optimize the reuse of $\mathcal{C}_{target}$ despite the personalized prefixes, next-generation RAG serving frameworks (e.g., LMCache \cite{liu2025lmcache,yao2025cacheblend}) utilize \textit{non-prefix KV cache fusion}. These frameworks dynamically locate and reuse cached chunk tensors even when they appear at arbitrary, non-prefix positions. 

\noindent\textbf{Micro-Architectural Fallback: Why Traditional Attacks Fail.} 
This architectural evolution completely blinds the traditional timing oracle. In a realistic RAG scenario, an attacker might correctly guess the retrieved document but will inevitably fail to perfectly guess the victim's unique, hidden system prefix. Even a single-token deviation at the root of the prompt breaks the strict prefix alignment. As illustrated in the execution timeline in Figure~\ref{fig:timeline_llm_rag}, standard caching mechanisms will aggressively isolate the request upon detecting this mismatch at the first token of the prefix. Consequently, the engine treats the request as a complete Cache Miss, falling back to a computationally expensive Full Attention Prefill for the entire sequence. When both correct and incorrect document guesses result in an identical full-prefill pathway, the timing difference ($\Delta T$) collapses to zero, annihilating the side-channel.

\begin{figure}[t] 
  \centering
  
  \begin{subfigure}[b]{0.24\textwidth}
    \centering
    \includegraphics[width=\linewidth]{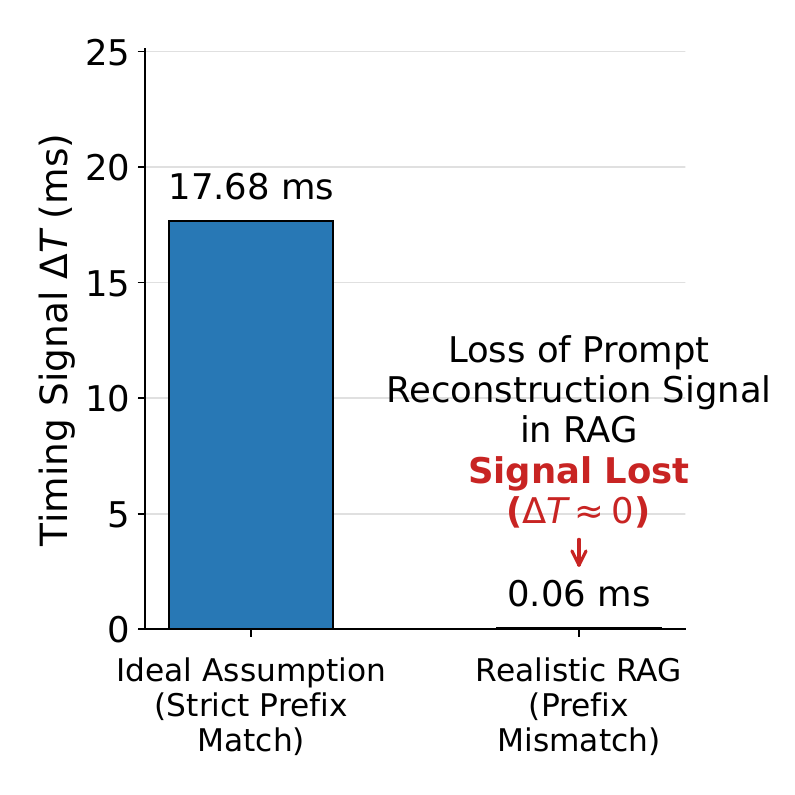}
    \caption{Macroscopic timing signal collapse ($\Delta T$).}
    \label{fig:motivation_bar}
  \end{subfigure}
  \hfill 
  \begin{subfigure}[b]{0.24\textwidth}
    \centering
    \includegraphics[width=\linewidth]{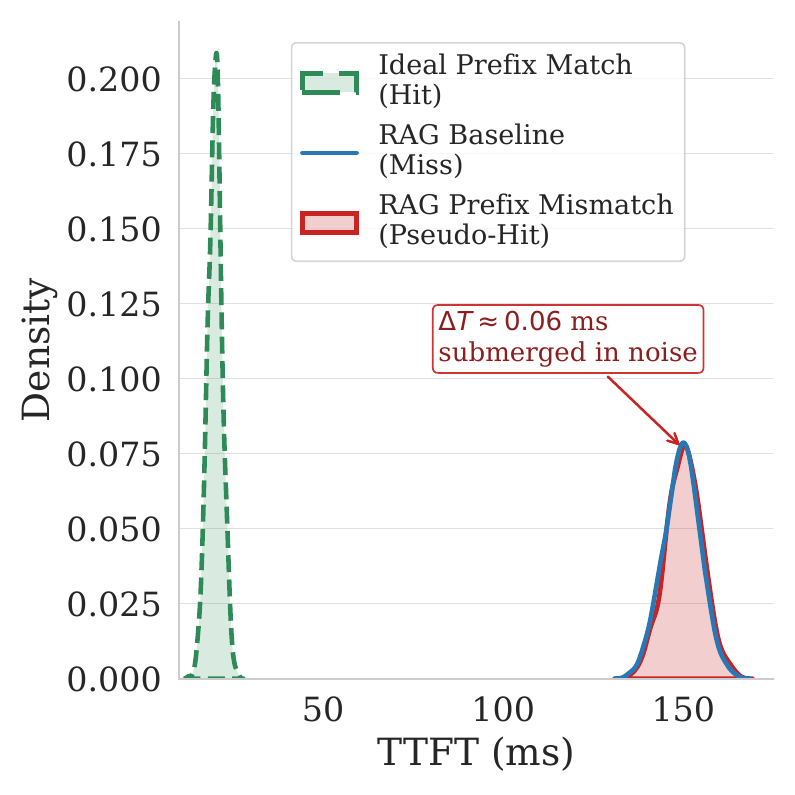}
    \caption{Statistical submergence of $\Delta T$ in system noise.}
    \label{fig:motivation_density}
  \end{subfigure}
  
  \caption{Vulnerability of traditional timing side-channel attacks in multi-tenant RAG frameworks. (a) In an ideal strict prefix matching scenario, a successful guess yields an exploitable timing signal ($\Delta T \approx 17.68$ ms). In realistic RAG environments, the inevitable prefix mismatch triggers full re-computation, causing the macroscopic signal to collapse. (b) The resulting marginal timing difference ($\Delta T \approx 0.06$ ms) is statistically submerged within system noise, rendering Cache Miss and Pseudo-Hit practically indistinguishable.}
  \label{fig:motivation_failure_combined}
\end{figure}

\noindent\textbf{Empirical Validation.} 
To empirically validate this signal collapse, we conducted a controlled side-channel experiment using the LLaMA-2-7B model. We simulated a prompt reconstruction attack targeting a 300-token document under two strictly controlled scenarios using vLLM \cite{kwon2023efficient} and LMCache \cite{liu2025lmcache,yao2025cacheblend}:
\begin{itemize}
    \item \textit{Ideal Assumption (Strict Prefix Match):} The attacker perfectly guesses both the system prefix and the target document, representing the optimistic threat model of prior works.
    \item \textit{Realistic RAG (Prefix Mismatch):} The attacker targets the correct document but uses an incorrect system prefix. To rigorously isolate the computational overhead and eliminate length bias, we forced the tokenizer to align the absolute lengths of both the correct and incorrect probes to 300 tokens.
\end{itemize}

As shown in Figure~\ref{fig:motivation_failure_combined}(a), the results reveal a stark contrast. Under the ideal assumption, a cache hit yields a substantial and easily exploitable timing signal ($\Delta T \approx 17.68$ ms). However, in the realistic RAG scenario, the minor prefix mismatch forces a full prefill, crushing the timing signal to a mere $\Delta T \approx 0.06$ ms. Furthermore, as depicted in the Probability Density Function (KDE) plot in Figure~\ref{fig:motivation_failure_combined}(b), this 0.06 ms residual difference is entirely swallowed by the systemic noise of kernel scheduling and network jitter. The distributions for a Cache Miss and a ``Pseudo-Hit'' perfectly overlap, proving that existing prefix-dependent side-channel attacks are statistically and fundamentally ineffective in RAG workloads.

\noindent\textbf{Orthogonality to Existing RAG Security Research.} 
Recent security research on RAG systems has primarily focused on application-layer threats and knowledge base vulnerabilities. A significant line of work investigates data poisoning and adversarial manipulation (e.g., PoisonedRAG \cite{zou2025poisonedrag}, FlippedRAG \cite{chen2025flippedrag}, Topic-FlipRAG \cite{gong2025topic}), demonstrating how malicious documents can manipulate LLM outputs. Similarly, works like ImportSnare \cite{ye2025importsnare} and CodeGuarder \cite{lin2025give} address code vulnerabilities injected through untrusted retrieval manuals. Other prominent works focus on availability via ``jamming'' attacks \cite{shafran2025machine} or Intellectual Property (IP) protection via watermarking \cite{lv2025rag}. 

While these pioneering works comprehensively address \textit{content-level} security originating from malicious inputs or untrusted corpora, they implicitly assume that the underlying cloud serving infrastructure securely isolates concurrent users. They do not investigate the \textit{system-level} and \textit{micro-architectural} vulnerabilities introduced by multi-tenant performance optimizations. These studies are orthogonal to the timing side-channels caused by continuous batching and non-prefix KV cache fusion. Consequently, their threat models cannot address the structural privacy leaks targeted in this paper, where an unprivileged attacker extracts sensitive prompt properties merely by exploiting the memory blending dynamics of advanced RAG serving engines.

\section{Overview}
\label{sec:overview}

Retrieval-Augmented Generation serving has become ubiquitous, powering a broad spectrum of applications ranging from large-scale, multi-tenant public cloud platforms to internal enterprise intelligent assistants. However, this widespread deployment introduces severe, yet previously unexplored, privacy challenges. To the best of our knowledge, this is the first work to systematically expose and exploit micro-architectural side-channel vulnerabilities inherent in modern RAG serving pipelines. In this section, we establish the foundational framework for our study. We first formulate a unified system model that characterizes the modular architecture of advanced RAG serving engines (Section~\ref{subsec:system_model}). Next, we rigorously define the threat model, detailing the attacker's capabilities, assumptions, and privacy-extraction goals (Section~\ref{subsec:threat_model}). Subsequently, we analyze the emerging micro-architectural side-channel opportunities and provide a high-level overview of our proposed attack methodology (Section~\ref{subsec:attack_overview}). Finally, recognizing the critical need for remediation, we formalize a comprehensive defense policy designed to systematically mitigate these vulnerabilities while preserving serving efficiency (Section~\ref{subsec:defense_policy}).

\subsection{System Model}
\label{subsec:system_model}

\begin{figure}[t]
  \centering
  \includegraphics[width=0.95\linewidth]{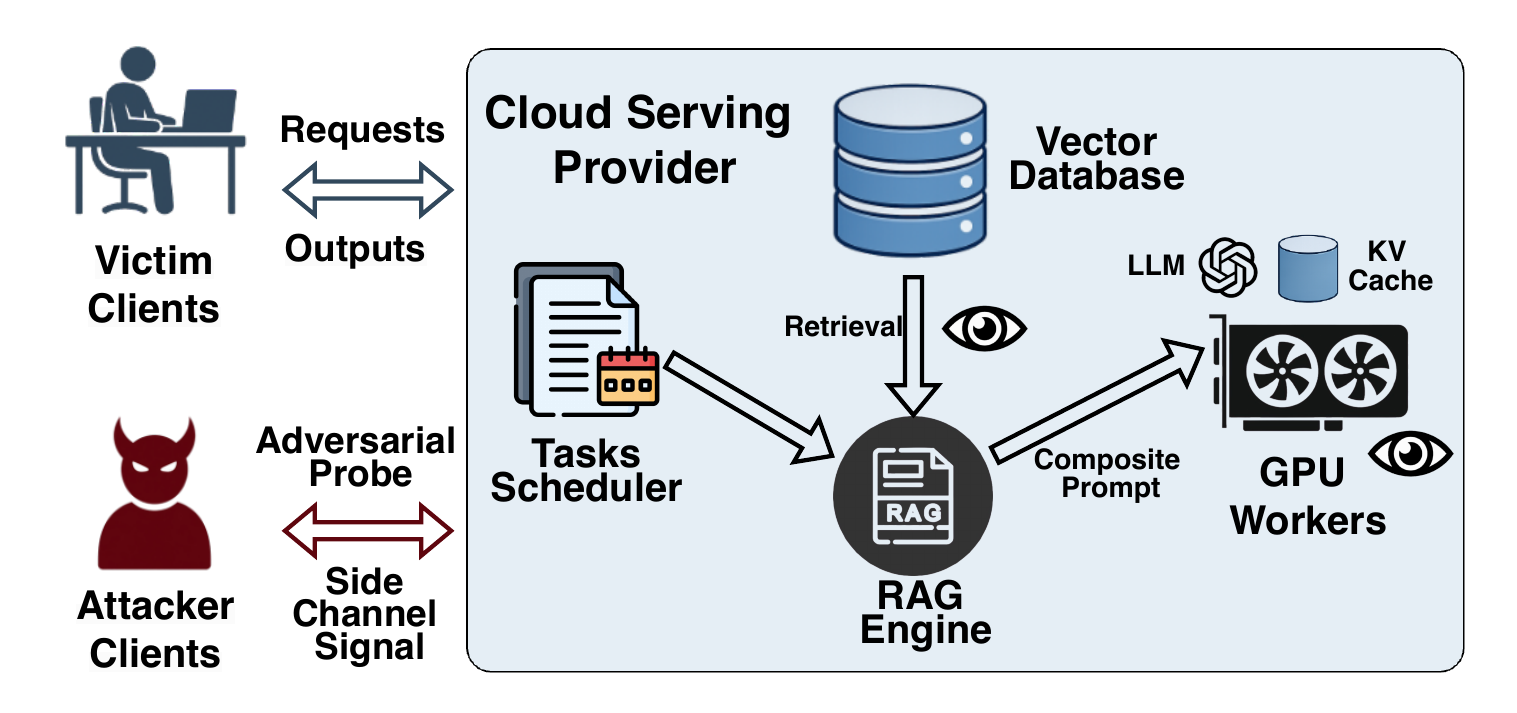} 
  \caption{High-level overview of our system model for multi-tenant RAG serving.}
  \label{fig:system_model}
  \vspace{-0.1in}
\end{figure}

Our research targets modern multi-tenant LLM serving engines (denoted as $\mathcal{E}_{sys}$) optimized for RAG workloads. To mitigate the massive I/O and computation bottlenecks caused by long-context documents, advanced frameworks (e.g., LMCache \cite{liu2025lmcache,yao2025cacheblend}) implement a \textit{Non-Prefix KV Cache Fusion} mechanism. As illustrated in Figure~\ref{fig:system_model}, our system architecture comprises a retrieval layer and a serving layer interconnected within a cloud infrastructure, enabling arbitrary, position-independent reuse of granular knowledge chunks. 

\noindent\textbf{Entities.} We identify three principal entities within this multi-tenant ecosystem:

\begin{itemize}
    \item \textbf{Cloud Serving Provider.} The encompassing infrastructure that hosts $\mathcal{E}_{sys}$. As depicted in Figure~\ref{fig:system_model}, it integrates four core internal components: the \textit{Tasks Scheduler} that queues incoming requests, the \textit{Vector Database} storing retrievable documents, the \textit{RAG Engine} acting as the central orchestrator, and the \textit{GPU Workers} equipped with the \textit{LLM} and a global, unified \textit{KV Cache} pool (denoted as $\Omega_{shared}$). Upon receiving a task, the RAG Engine fetches relevant document chunks via \textit{Retrieval}, assembles a \textit{Composite Prompt}, and dispatches it to the GPU Workers where pre-computed tensors are dynamically loaded from $\Omega_{shared}$ and fused with new inputs.
    
    \item \textbf{Victim Clients.} Legitimate users interacting with the RAG service via standard \textit{Requests} and \textit{Outputs}. The victim's input sequence, denoted as $X_{vic}$, is structurally composed of three segments: $X_{vic} = \mathcal{P}_{vic} \parallel \mathcal{C}_{target} \parallel \mathcal{Q}_{vic}$, where $\parallel$ denotes sequence concatenation. Here, $\mathcal{P}_{vic}$ represents the victim's highly confidential private prefix. To illustrate, in a healthcare RAG application, $\mathcal{P}_{vic}$ typically encapsulates highly sensitive implicit context dynamically injected by the backend (e.g., a hidden system profile detailing the user's medical history, such as \textit{``Patient is HIV-positive and currently on antiretroviral therapy''}). $\mathcal{C}_{target}$ represents the shared public knowledge chunk retrieved from the vector database (e.g., \textit{the pharmacological contraindications of a common antibiotic}), and $\mathcal{Q}_{vic}$ is the explicit, often benign, user query (e.g., \textit{``Are there any severe side effects if I take this medication?''}). By system design, both the semantic content and the structural length ($N_{vic} = |\mathcal{P}_{vic}|$) of the private prefix are strictly isolated from other tenants, even when $\mathcal{P}_{vic}$ is semantically intertwined with the shared retrieved chunk $\mathcal{C}_{target}$.
    
    \item \textbf{Attacker Clients.} A co-located adversarial tenant utilizing the same $\mathcal{E}_{sys}$ instance. The attacker submits an \textit{Adversarial Probe}, structurally crafted as $X_{adv} = \mathcal{P}_{adv} \parallel \mathcal{C}_{target} \parallel \mathcal{Q}_{adv}$. Here, $\mathcal{P}_{adv}$ is an attacker-controlled prefix with a meticulously chosen length ($N_{adv} = |\mathcal{P}_{adv}|$). Without requiring any special system privileges, the attacker's objective is to systematically modulate $N_{adv}$ to trigger and monitor deterministic \textit{Side-Channel Signals} (e.g., latency anomalies) during the engine's cross-chunk KV fusion phase. By analyzing these signals, the attacker dynamically extracts the private properties—both the structural length $N_{vic}$ and the deep semantic content (e.g., the hidden HIV status) of $\mathcal{P}_{vic}$.

\end{itemize}

\noindent\textbf{Policy Specifications.} The operations within our system model are governed by a suite of policies tailored for advanced RAG serving:

\begin{itemize}
    \item \textbf{Retrieval Policy ($\mathcal{P}_{ret}$).} When a client submits a query, the RAG Engine queries the Vector Database to fetch the top-$k$ most semantically relevant document chunks. Crucially, these chunks are treated as modular components that can be placed at arbitrary positions within the final composite prompt, independent of the sequence's starting tokens.
    
    \item \textbf{KV Cache Policy ($\mathcal{P}_{KV}$).} Unlike traditional strict prefix caching that evicts cache entries upon the first token mismatch, $\mathcal{E}_{sys}$ maintains $\Omega_{shared}$ within the GPU Workers as a modular memory pool. The KV states of frequently retrieved public chunks ($\mathcal{C}_{target}$) are persistently stored. Access to $\Omega_{shared}$ is shared globally across all tenants (clients) to maximize memory efficiency.
    
    \item \textbf{Chunk-Aware Blending Policy ($\mathcal{P}_{blend}$).} While advanced engines utilize \textit{KV Blending} to stitch the pre-computed tensors of $\mathcal{C}_{target}$ with the private prefix ($\mathcal{P}_{vic}$ or $\mathcal{P}_{adv}$), the critical operational unit of this policy is the fixed-size memory chunk (e.g., the block size $B=256$ tokens). During the assembly of $X_{vic}$ or $X_{adv}$, the GPU Workers selectively route and load the maximally aligned \textit{full chunks} of $\mathcal{C}_{target}$ directly from $\Omega_{shared}$. Consequently, any unaligned residual tokens—referred to as the sequence \textit{tail}—cannot be directly memory-mapped and must undergo \textit{Selective Re-computation} to restore attention states. As we will demonstrate, this chunk-level memory allocation combined with the tail re-computation introduces a deterministic, step-wave-like execution latency, serving as the primary side-channel leakage source continuously monitored by the attacker.
\end{itemize}

\subsection{Threat Model}
\label{subsec:threat_model}

We consider a realistic and highly restrictive threat model that strictly adheres to the operational constraints of modern cloud-based LLM-as-a-Service (LLMaaS) environments. Crucially, the adversary operates entirely in a black-box setting from a system administration perspective.

\noindent\textbf{Attacker's Goal.} The ultimate objective of the adversary is to breach the KV cache memory isolation of the RAG serving engine, systematically reconstructing the highly confidential private prefix ($\mathcal{P}_{vic}$) of co-located tenants. Depending on the adversary's background knowledge of the LLM application (e.g., whether the adversary and the victim are using the same service template), the specific extraction target may vary from a single sensitive user variable to the entire proprietary system instruction. We examine these diverse extraction scenarios in detail in Section~\ref{sec:evaluation}. To achieve this, the attacker executes a two-phase strategy:
\begin{enumerate}
    \item \textbf{Phase I: Structural Fingerprinting.} Exploiting micro-architectural timing anomalies (e.g., the \textit{Step-Wave Effect}) to mathematically deduce the exact hidden token length ($N_{vic}$) of the victim's context.
    \item \textbf{Phase II: Semantic Extraction.} Utilizing $N_{vic}$ as an indispensable anchor to manipulate boundary collisions, forcing the engine to leak the exact semantic content of $\mathcal{P}_{vic}$ token-by-token.
\end{enumerate}

\noindent\textbf{Attacker's Capabilities.} The adversary possesses the same capabilities as an ordinary user, with no direct control or visibility over the inference server's physical hardware, hypervisor, or internal memory states (e.g., KV cache pointers). They can only interact with the server through standard client APIs to submit textual prompts ($X_{adv}$) and observe the end-to-end response latency (specifically, the TTFT). 

Following standard assumptions in prior side-channel research, we assume the attacker is aware of the default internal mechanisms of the LLM serving framework, specifically the deployment of chunk-aware non-prefix KV cache fusion. The background knowledge extends to the target LLM's public tokenizer, which is widely accessible for both open-source and commercial models. Furthermore, to execute Phase II efficiently, we assume the attacker possesses a standard offline auxiliary LLM. This local model is utilized to generate high-probability token candidates based on contextual semantics, drastically reducing the search space required for the online oracle.

\noindent\textbf{System Assumptions.} The attack relies on two practical preconditions: (1) \textit{Co-residency:} The attacker achieves logical co-residency with the victim on the same physical serving node, a probabilistic inevitability in dense multi-tenant architectures \cite{ristenpart2009hey,yu2022orca,wu2025know,wang2024gpu}. (2) \textit{Shared Corpus:} The attacker can identify or reasonably guess the public knowledge chunks ($\mathcal{C}_{target}$) retrieved by the RAG system. This is highly realistic in domain-specific applications (e.g., enterprise legal analysis) where multiple tenants frequently query a centralized, deterministic knowledge base \cite{liu2025lmcache,yao2025cacheblend}.

\begin{figure}[t]
    \centering
    \includegraphics[width=0.5\textwidth]{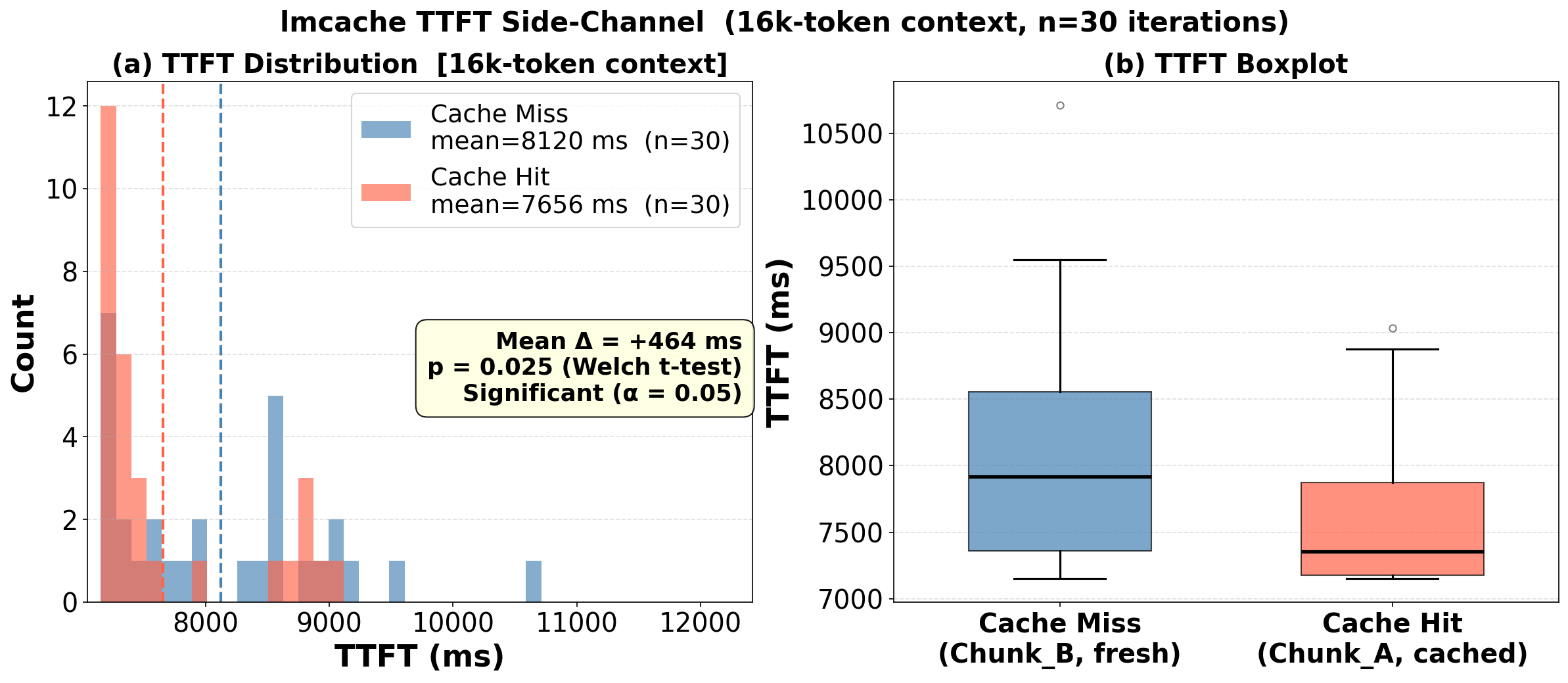}
    \caption{TTFT side-channel in a 16K-token RAG scenario ($n=30$). (a) Histogram showing a distinct latency shift for cache hits. (b) Boxplot confirming a statistically significant latency drop ($\Delta \approx 464$ ms, Welch's $p=0.025$) upon retrieving a cached document. This validates that cross-tenant timing leakage pierces systemic noise.}
    \label{fig:rag_ttft_attack}
    \vspace{-0.1in}
\end{figure}

\subsection{Attack Overview}
\label{subsec:attack_overview}

To breach the KV cache memory isolation enforced by modern RAG serving frameworks, we propose a novel two-phase attack methodology. This attack exploits deterministic micro-architectural timing anomalies—introduced during selective KV cache re-computation and cross-chunk fusion—to systematically compromise both access privacy and semantic confidentiality.

\noindent\textbf{Phase I: Structural Fingerprinting via Hierarchical TTFT Side-Channels.} 
The first phase of our attack establishes a high-resolution structural blueprint of the victim's hidden context by exploiting hierarchical timing leakages. When the serving engine ($\mathcal{E}_{sys}$) fuses a shared, pre-computed KV tensor ($\mathcal{C}_{target}$) into an attacker's context, it circumvents the computationally expensive full prefill phase. We demonstrate that this micro-architectural acceleration can be observed and exploited at two distinct resolutions:
\begin{figure*}[t]
    \centering
    \includegraphics[width=\textwidth]{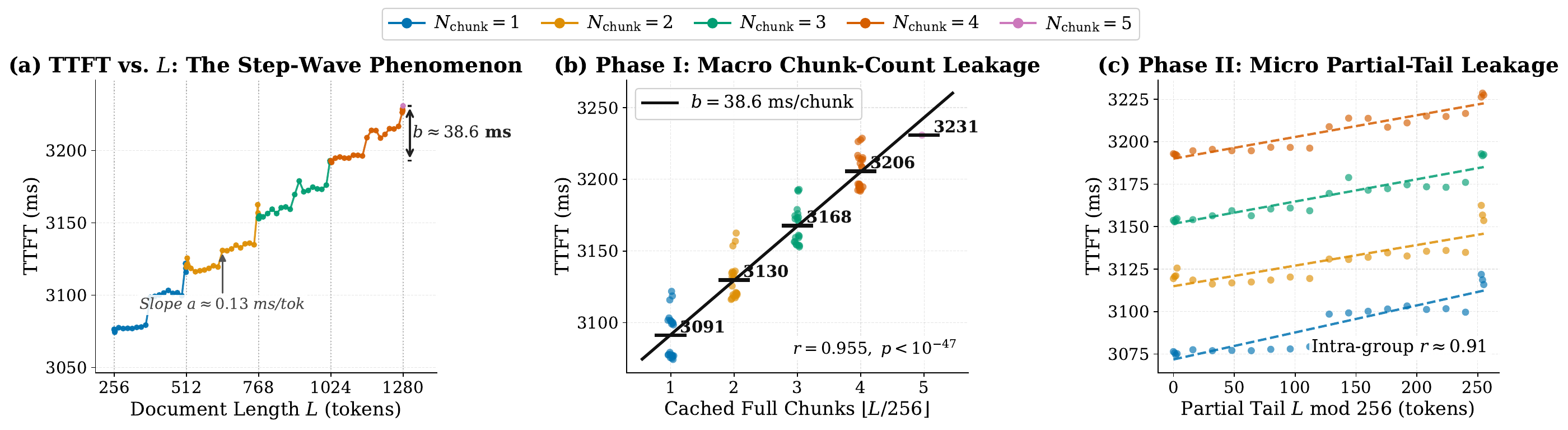}
    \vspace{-0.15in} 
    \caption{Micro-architectural profiling of the Step-Wave TTFT side-channel in LMCache. 
    (a) The overall $\mathcal{T}_{TTFT}$ exhibits a distinctive step-wave pattern as document length $L$ increases, defying the theoretical sawtooth model. 
    (b) Phase I: The absolute $\mathcal{T}_{TTFT}$ strictly correlates with the number of full chunks ($\lfloor L/256 \rfloor$), exposing the macro-level length with zero error ($b \approx 38.6$ ms/chunk). 
    (c) Phase II: Within each chunk group, the residual $\mathcal{T}_{TTFT}$ linearly correlates with the partial tail length ($L \bmod 256$), forming a micro-level length oracle with an intra-group correlation of $r \approx 0.91$ ($a \approx 0.13$ ms/tok).}
    \label{fig:step_wave}
    \vspace{-0.1in} 
\end{figure*}

\textit{1) The Binary Cache-Hit Oracle:} As empirically validated in Figure~\ref{fig:rag_ttft_attack}, profiling the TTFT ($\mathcal{T}_{TTFT}$) across a massive 16K-token RAG context reveals a pronounced distribution gap. A fresh cache miss incurs a substantial prefill latency (mean $\mathcal{T}_{TTFT} = 8120$ ms), whereas a cache hit leveraging the fusion mechanism significantly accelerates the response (mean $\mathcal{T}_{TTFT} = 7656$ ms). This latency reduction ($\Delta \approx 464$ ms) is highly statistically significant (e.g., Welch's t-test \cite{welch1938significance}, $p = 0.025$ at the significance level $\alpha = 0.05,\ n=30$). This formally proves that the observed timing discrepancy is not an artifact of random system noise or network jitter, but a deterministic binary signal. By crafting probe queries that trigger the retrieval of specific sensitive chunks, an attacker can exploit this oracle to definitively sniff the victim's retrieval intent.

\textit{2) The Step-Wave Length Oracle:} The vulnerability extends far beyond simple binary detection. As demonstrated in our micro-architectural profiling (Figure~\ref{fig:step_wave}), the underlying fusion mechanism inadvertently acts as a high-resolution length oracle. As illustrated in Figure~\ref{fig:step_wave}(a), rather than following a theoretical sawtooth pattern, $\mathcal{T}_{TTFT}$ exhibits a deterministic \textit{Step-Wave} distribution as document length $L$ increases. This phenomenon is driven by the fixed-size block management of the engine (where $B$ denotes the block size, e.g., $B=256$ in frameworks like lmcache \cite{liu2025lmcache,yao2025cacheblend}). 

This empowers the attacker to hierarchically extract the structural fingerprint: First, as shown in Figure~\ref{fig:step_wave}(b), the absolute $\mathcal{T}_{TTFT}$ strictly correlates with the number of full chunks ($\lfloor L/B \rfloor$), exposing the macro-level length with zero error due to the per-chunk blending overhead ($b \approx 38.6$ ms/chunk). Second, as shown in Figure~\ref{fig:step_wave}(c), within each chunk group, the residual latency linearly exposes the partial tail length ($L \bmod B$) driven by the tail prefill overhead ($a \approx 0.13$ ms/tok, intra-group $r \approx 0.91$). By synthesizing these signals, the attacker can pinpoint the precise token length of the hidden context ($N_{vic}$) with an unprecedented accuracy of $\pm 30$ tokens.

\noindent\textbf{Phase II: Semantic Extraction via SpliceLeak.} 
Armed with the high-resolution structural blueprint derived from the step-wave oracle, the attacker transitions from structural fingerprinting to targeted semantic extraction. Crucially, uncovering the precise token length ($N_{vic}$) resolves the previously intractable challenge of block misalignment. It empowers the attacker to strategically craft probing prefixes that perfectly align with the engine's fixed KV block boundaries (e.g., $B=256$-token chunks in \textit{CacheBlend}). By precisely manipulating the boundary tokens immediately adjacent to the fusion point of $\mathcal{C}_{target}$, and exploiting the deterministic KV block allocation and selective re-computation logic of the engine, the attacker can force the system to systematically leak the actual semantic contents comprising the victim's prompt ($\mathcal{P}_{vic}$). The rigorous mathematical modeling of this boundary-aligned leakage channel, alongside the end-to-end extraction algorithms, will be comprehensively detailed in Section~\ref{sec:spliceleak}.

\subsection{Defense Policy}
\label{subsec:defense_policy}

The discovery of the SpliceLeak vulnerability underscores a critical tension between aggressive performance optimization (e.g., arbitrary position chunk sharing) and fundamental KV cache memory isolation in multi-tenant LLM serving. Recognizing the urgent need for remediation, we propose a comprehensive defense policy designed to systematically mitigate these micro-architectural side-channels while preserving the throughput benefits of RAG-optimized caching. 

Our proposed mitigation strategies operate across two distinct architectural layers:
\begin{itemize}
    \item \textbf{System-Level Isolation (Namespace Partitioning):} We advocate for the enforcement of strict tenant-level isolation within the global KV cache pool ($\Omega_{shared}$). By configuring the serving engine (e.g., vLLM or LMCache) to bind cached chunks to specific tenant cryptographic namespaces, we physically prevent cross-tenant non-prefix fusion. While this reduces the global cache hit rate for public documents, it decisively neutralizes the attacker's ability to trigger the selective re-computation penalty on a victim's cached tensor.
    \item \textbf{Micro-Architectural Obfuscation (Alignment Noise Injection):} For environments where cross-tenant sharing is strictly necessary for economic viability, we propose modifying the underlying PagedAttention scheduler. By intentionally injecting randomized padding tokens (dummy computations) during the fusion phase, the engine can artificially misalign the blocks or equalize the selective re-computation latency. This effectively destroys the high-resolution timing profile required for the Phase I Structural Fingerprinting attack, submerging the alignment signal beneath an impenetrable layer of algorithmic noise.
\end{itemize}
The detailed implementation and performance overhead evaluation of these defense mechanisms are presented in Section~\ref{sec:splicedefense}.

\section{SpliceLeak: Semantic Extraction Methodology}
\label{sec:spliceleak}

In Phase I of our methodology, we established that the micro-architectural \textit{Step-Wave Effect} reliably exposes the exact structural length ($N_{vic}$) of the victim's hidden private prefix $\mathcal{P}_{vic}$. In this section, we detail \textbf{SpliceLeak}, the second phase of our attack. We first formalize the physical boundary-anchored token extraction mechanism, and subsequently introduce an \textit{LLM-driven Agentic framework} that synergizes micro-architectural feedback with advanced contextual reasoning to execute scenario-driven privacy extraction.

\subsection{Boundary-Anchored Extraction Mechanism}
\label{subsec:extraction_mechanism}

The core vulnerability enabling SpliceLeak's semantic extraction lies in the engine's deterministic Key-Value (KV) deviation evaluation during cross-chunk fusion. Let the victim's targeted private prefix be $\mathcal{P}_{vic} = [t_1, t_2, \dots, t_{N_{vic}}]$. The attacker initially targets the boundary token $t_{N_{vic}}$, positioned immediately adjacent to the shared retrieved chunk $\mathcal{C}_{target}$.

Armed with $N_{vic}$, the attacker constructs an adversarial probe $X_{adv}$ composed of a precisely calculated padding sequence $\mathcal{P}_{pad}$ (where $|\mathcal{P}_{pad}| = N_{vic} - 1$), a guessed candidate token $t_{guess} \in \mathcal{V}$ (where $\mathcal{V}$ is the model's vocabulary), and the shared chunk $\mathcal{C}_{target}$:
\begin{equation}
    X_{adv} = \mathcal{P}_{pad} \parallel t_{guess} \parallel \mathcal{C}_{target}
\end{equation}

When the serving engine processes $X_{adv}$, it evaluates the L2-deviation between the incoming KV states and the cached states of $\mathcal{C}_{target}$. If the guessed token exactly matches the victim's true boundary token ($t_{guess} = t_{N_{vic}}$), semantic continuity is preserved, minimizing the cross-attention discrepancy. This triggers a zero-collision state, drastically reducing the selective re-computation budget. The attacker observes this as a deterministic, statistically significant drop in $\mathcal{T}_{TTFT}$. By shifting the padding length leftward iteratively and appending successfully extracted tokens, the attacker can traverse the prefix in reverse.

\subsection{LLM-Driven Agentic Extraction Framework}
\label{subsec:agentic_extraction}

While a naive right-to-left reverse brute-force across the entire LLM vocabulary $\mathcal{V}$ (typically $>32,000$ tokens) is theoretically possible, it incurs a prohibitively high API request cost ($\text{Req./tok}$). To bridge the gap between a theoretical hardware vulnerability and a practical, lethal exploit, we formulate the SpliceLeak attacker as a \textbf{Side-Channel Feedback-Driven Extraction Agent} ($\mathcal{A}_{LLM}$). 

\textbf{Agentic Capabilities:} Depending on the attacker's operational constraints, $\mathcal{A}_{LLM}$ can be instantiated in two modalities. In an extremely restrictive, air-gapped threat model, the attacker can deploy a lightweight local open-source model (e.g., Llama-3-8B \cite{grattafiori2024llama}) to ensure zero external reliance. Conversely, an unconstrained attacker can leverage state-of-the-art commercial APIs (e.g., GPT-5 \cite{singh2025openai} or Gemini 3 Pro \cite{google2025gemini3pro}). By feeding the known public chunk ($\mathcal{C}_{target}$) and any previously extracted tokens into $\mathcal{A}_{LLM}$, the Agent possesses immense contextual inference capabilities, generating highly constrained and accurate candidate subsets ($\mathcal{V}_{sub}$).

As formalized in Algorithm~\ref{alg:spliceleak}, this Agent operates in a continuous loop: reasoning about the context to propose $\mathcal{V}_{sub}$, acting by sending probes to the target engine, and using the physical $\mathcal{T}_{TTFT}$ side-channel dip as an absolute ground-truth perception oracle to validate its hypotheses.

\begin{algorithm}[t]
\caption{SpliceLeak: Agentic Right-to-Left Token Extraction}
\label{alg:spliceleak}
\begin{algorithmic}[1]
\Require Shared RAG chunk $\mathcal{C}_{target}$, Target prefix length $N_{vic}$
\Require Extraction Agent $\mathcal{A}_{LLM}$ (Local Model or External API)
\Require Threshold $\delta_{th}$ for TTFT deviation drop
\Ensure Extracted sequence $\mathcal{S}_{extracted}$
\State $\mathcal{S}_{extracted} \gets []$
\For{$i = N_{vic}$ \textbf{down to} $1$}
    \State $\mathcal{P}_{pad} \gets \text{GeneratePaddingSequence}(i - 1)$
    \State $\mathcal{V}_{sub} \gets \mathcal{A}_{LLM}\text{.PredictCandidates}(\mathcal{S}_{extracted}, \mathcal{C}_{target})$
    \State $\mathcal{T}_{baseline} \gets \infty$
    \State $t_{target} \gets \text{NULL}$
    \For{\textbf{each} $t_{guess} \in \mathcal{V}_{sub}$}
        \State $X_{adv} \gets \mathcal{P}_{pad} \parallel t_{guess} \parallel \mathcal{S}_{extracted} \parallel \mathcal{C}_{target}$
        \State $\mathcal{T}_{current} \gets \text{Measure\_TTFT}(X_{adv})$ \Comment{Side-Channel Oracle}
        \If{$\mathcal{T}_{current} < \mathcal{T}_{baseline} - \delta_{th}$}
            \State $\mathcal{T}_{baseline} \gets \mathcal{T}_{current}$
            \State $t_{target} \gets t_{guess}$
        \EndIf
    \EndFor
    \If{$t_{target} \neq \text{NULL}$}
        \State $\text{Prepend } t_{target} \text{ to } \mathcal{S}_{extracted}$
    \Else
        \State \textbf{break} \Comment{Agent prediction failed; halt extraction or fallback to beam-search}
    \EndIf
\EndFor
\State \Return $\mathcal{S}_{extracted}$
\end{algorithmic}
\end{algorithm}

\subsection{Execution across Threat Scenarios}
\label{subsec:attack_scenarios}

Empowered by this Agentic framework, SpliceLeak gracefully adapts across the three defined threat scenarios (Section~\ref{subsec:threat_model}), minimizing the search space and attack cost:

\begin{itemize}
    \item \textbf{Scenario A (Confidential Input Extraction):} When the rigid prompt template is known (e.g., \textit{``Patient profile: [STATUS]''}), the Agent $\mathcal{A}_{LLM}$ utilizes this structural constraint to generate extremely targeted candidates for the private variables (e.g., \textit{``HIV'', ``Diabetic''}). This drastically shrinks the search space to $|\mathcal{V}_{sub}| < 10$, extracting critical personal secrets with a negligible probing budget.
    
    \item \textbf{Scenario B (Proprietary Template Reverse-Engineering):} When commercial RAG platforms shield their core system prompts, the Agent utilizes the known RAG document $\mathcal{C}_{target}$ to guide its predictions. By narrowing $\mathcal{V}_{sub}$ to structural instructions, role-playing personas, and connective grammar, the Agent iteratively strips the application of its proprietary intellectual property.
    
    \item \textbf{Scenario C (Zero-Knowledge Privacy Reconstruction):} Even in a complete zero-knowledge baseline lacking both template and user input constraints, the Agentic framework remains devastatingly effective. By continuously querying its internal language distribution to guess the preceding context of $\mathcal{C}_{target}$, the Agent treats the micro-architectural timing dip as an ultimate truth mechanism. It persistently walks backward, deterministically extracting deeply personal, free-form user queries token-by-token.
\end{itemize}

\section{SpliceDefense: Mitigating Non-Prefix Side-Channels}
\label{sec:splicedefense}

The discovery of the SpliceLeak vulnerability highlights a fundamental and previously unresolved tension in multi-tenant LLM serving: the conflict between aggressive memory deduplication (e.g., non-prefix chunk fusion) and strict micro-architectural isolation. To secure RAG serving frameworks without sacrificing the substantial throughput benefits of global KV cache sharing, we propose \textbf{SpliceDefense}, a synergistic, two-stage mitigation framework operating directly at the memory scheduling and attention computation layers.

\subsection{Defense I: Quantized Chunk Padding (QCP)}
\label{subsec:defense_qcp}

The root cause of SpliceLeak's Phase I structural fingerprinting is the periodic timing fluctuation that occurs when an attacker sweeps their prefix length across the engine's physical PagedAttention block boundaries. Rather than isolating cache namespaces—which completely destroys cross-tenant deduplication—we propose Quantized Chunk Padding (QCP) to eliminate the structural timing oracle.

\noindent\textbf{Mechanism.} QCP operates at the prompt processing phase. Before the engine allocates KV cache blocks for a tenant's private prefix $\mathcal{P}_{vic}$, the scheduler mathematically rounds the sequence length up to the nearest strict multiple of the block size $B$ (e.g., 256 tokens). The engine dynamically appends deterministic null-tokens to the private prefix such that $|\mathcal{P}'_{vic}| \equiv 0 \pmod B$. Consequently, any subsequently retrieved public chunk $\mathcal{C}_{target}$ is guaranteed to be fused perfectly at a block boundary, regardless of the original prompt length.

\noindent\textbf{Security and Trade-offs.} QCP provides deterministic immunity against the Phase I length extraction attack. By forcing all private prefixes to align uniformly to the block grid, the observable step-wave timing pattern $\mathbb{T} = \{\mathcal{T}_{TTFT}(N_{adv})\}$ is entirely flattened. Crucially, QCP retains 100\% of the global KV cache sharing capabilities for $\mathcal{C}_{target}$, fulfilling the primary objective of frameworks like LMCache. The only overhead is the computational and memory footprint of padding at most $B-1$ dummy tokens per request, which introduces a negligible microsecond-level latency penalty.

\subsection{Defense II: Constant-Time Boundary Fusion (CTBF)}
\label{subsec:defense_ctbf}

While QCP neutralizes structural length leaks, the Phase II semantic extraction relies on the micro-architectural L2-deviation dip that occurs during a correct token guess at the fusion boundary. Injecting random noise to mask this dip is vulnerable to statistical averaging attacks. Therefore, we introduce Constant-Time Boundary Fusion (CTBF), enforcing strict timing guarantees at the attention layer.

\noindent\textbf{Mechanism.} Modern fusion frameworks attempt to selectively skip re-computation if the incoming tensor states closely match the cached states (low L2-deviation). CTBF modifies this greedy optimization. When splicing a shared chunk $\mathcal{C}_{target}$ behind a private prefix, the engine enforces a constant, worst-case selective re-computation budget for the boundary block, regardless of semantic collisions. Even if the attacker's guessed token $t_{guess}$ perfectly matches the victim's true token, the GPU kernel is artificially mandated to execute the exact same number of FLOPs as it would for a completely mismatched token.

\noindent\textbf{Security and Trade-offs.} By enforcing constant-time execution specifically at the cross-chunk fusion boundary, CTBF permanently destroys the timing asymmetry required for token-by-token extraction. The attacker observes a static $\mathcal{T}_{TTFT}$ regardless of their candidate guesses. The performance trade-off is exquisitely balanced: the system still avoids the massive full prefill of the entire $\mathcal{C}_{target}$ (which may span thousands of tokens), only sacrificing the micro-optimization of skipping the boundary block re-computation. This ensures that the system maintains near-optimal serving throughput while achieving strict micro-architectural side-channel resistance.

\section{Evaluation}
\label{sec:evaluation}

In this section, we comprehensively evaluate the effectiveness of SpliceLeak across diverse real-world scenarios and assess the security-performance trade-offs of SpliceDefense. Our evaluation is designed to answer the following core research questions: 
(1) How effective and costly is SpliceLeak in extracting private inputs, proprietary templates, and full prompts under varying levels of adversary knowledge? 
(2) Does the underlying attention architecture (MHA vs. GQA) inherently mitigate or exacerbate the side-channel vulnerability? 
(3) Can SpliceDefense effectively neutralize the structural leakage with an acceptable system overhead?

\subsection{Experimental Setup}
\label{subsec:exp_setup}

\noindent\textbf{Hardware and Software Environment.} All experiments are conducted on a local GPU cluster equipped with 4 NVIDIA A40 GPUs (48GB VRAM each) and 256GB of system RAM. For the serving backend, we utilize a state-of-the-art continuous batching inference engine, \texttt{vLLM} \cite{kwon2023efficient}, integrated with the open-source \texttt{LMCache} \cite{liu2025lmcache,yao2025cacheblend} framework to enable cross-request non-prefix KV cache sharing. 

\noindent\textbf{Target Models.} To demonstrate the architectural generalizability of the step-wave side-channel, we evaluate our attacks on two widely deployed LLM architectures. This allows us to observe if techniques like key-value head sharing inherently impact the attack surface.
\begin{itemize}
    \item LongChat-7B-v1.5-32K \cite{longchat,longchat7b}: A model utilizing the standard Multi-Head Attention (MHA) mechanism, widely used for long-context RAG applications.
    \item Qwen2.5-7B-Instruct \cite{hui2024qwen2}: A state-of-the-art model employing Grouped-Query Attention (GQA), representative of modern high-efficiency LLMs that significantly compress KV cache sizes.
\end{itemize}

\noindent\textbf{Simulated Multi-Tenant Environment.} 
To evaluate attack resilience, we simulate a multi-tenant LLM service adopting the workload methodology from PROMPTPEEK \cite{wu2025know}. Background traffic follows a Poisson distribution with a per-client arrival rate of $\lambda = 0.004$ req/s. Despite the seemingly low rate, the extensive memory footprint of long-context RAG documents effectively saturates the continuous batching scheduler (max batch size 16), inducing authentic queuing and scheduling jitter representative of production environments.

\noindent\textbf{Datasets and Scenarios.} To align with our defined threat models, we construct evaluation workloads using a combination of standard RAG context benchmarks (e.g., 2WikiMQA \cite{ho2020constructing} and Musique datasets \cite{trivedi2022musique}), as utilized in recent KV cache fusion evaluations like CacheBlend \cite{yao2025cacheblend}), prepended with specific prompt styles drawn from large-scale datasets. Following the scenario design in PROMPTPEEK \cite{wu2025know}, we utilize:
\begin{itemize}
    \item \textbf{Cloze-style Prompts (for Scenario A):} We sample prompts from PromptBase, where the rigid template is known, but specific placeholders (e.g., patient conditions, financial figures) represent the targeted private input.
    \item \textbf{Role/Instruction-based Prompts (for Scenario B):} We utilize the \textit{awesome-chatgpt-prompts} and \textit{alpaca-gpt4} datasets \cite{peng2023instruction} to simulate proprietary system instructions that the attacker aims to reverse-engineer.
    \item \textbf{General Prompts (for Scenario C):} We sample unstructured, multi-round dialogue data from the \textit{Ultrachat} dataset to evaluate the baseline whole-prompt reconstruction without any prior knowledge.
\end{itemize}

\noindent\textbf{Evaluation Metrics.} We rigorously quantify the attack efficacy and defense overhead using the following standard metrics similar to PROMPTPEEK \cite{wu2025know}:
\begin{itemize}
    \item \textbf{Success Rate (SR):} The percentage of targeted prompts that are fully or partially extracted successfully.
    \item \textbf{Reversal Ratio (RR):} The ratio of the extracted token length to the total hidden prompt length, measuring the completeness of the extraction.
    \item \textbf{Attack Cost (Req./tok \& Req./inp):} The average number of probing requests required to extract a single token and the entire targeted input, respectively. This quantifies the stealthiness and efficiency of SpliceLeak.
    \item \textbf{System Overhead:} The Time-To-First-Token (TTFT) latency penalty and throughput degradation introduced by SpliceDefense compared to the vanilla LMCache engine.
\end{itemize}

\noindent\textbf{Extraction Agent ($\mathcal{A}_{LLM}$) Setup.} 
To instantiate $\mathcal{A}_{LLM}$, we assume an attacker utilizing state-of-the-art commercial LLM APIs. This reflects the \textit{asymmetrical technical advantage} of modern threat models, where powerful cloud reasoning is available at a negligible cost. Consequently, for Scenarios B and C, we power $\mathcal{A}_{LLM}$ with the \texttt{Gemini-3-Pro} API. To demonstrate the out-of-the-box lethality of SpliceLeak, we strictly avoid any fine-tuning. The API is configured with a temperature of $0.7$ and top-$p$ of $0.9$, optimally balancing deterministic structural comprehension of $\mathcal{C}_{target}$ with sufficient lexical diversity to generate high-coverage candidate subsets ($|\mathcal{V}_{sub}| = 20$).

\subsection{Effectiveness and Attack Cost}
\label{subsec:eval_effectiveness}

We first evaluate the efficacy of SpliceLeak across the three defined threat scenarios, measuring extraction accuracy, required API request costs, and the underlying architectural impacts.

\begin{figure}[t] 
    \centering
    \begin{subfigure}[b]{0.24\textwidth}
        \centering
        \includegraphics[width=\textwidth]{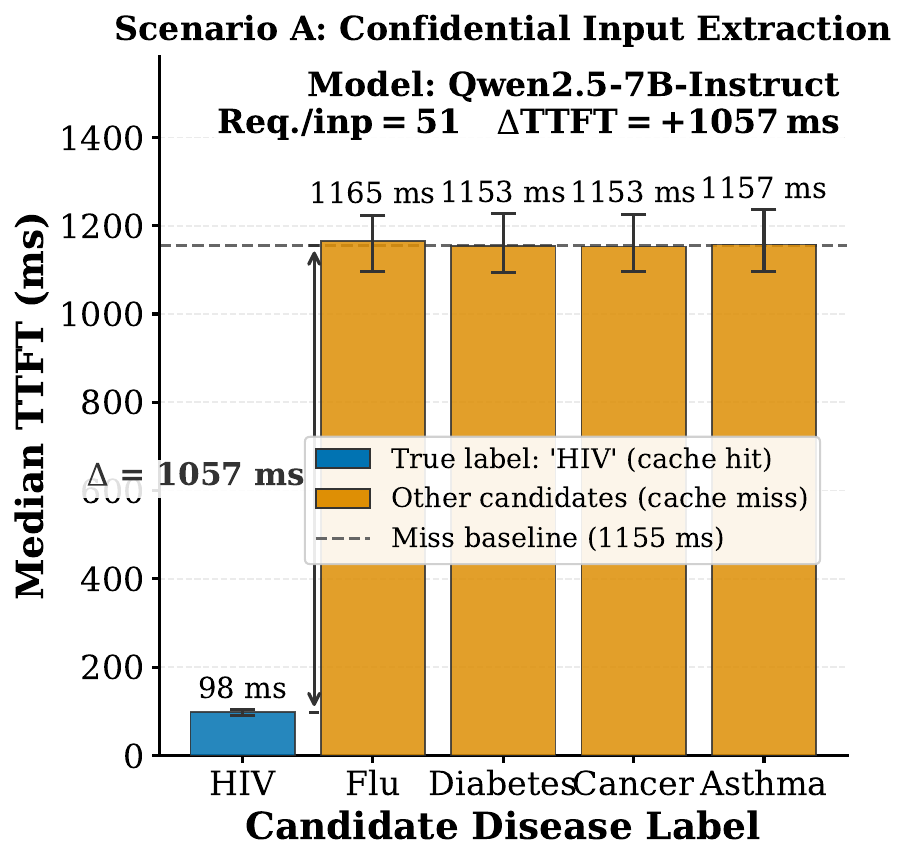}
        \caption{Micro-benchmark: Single targeted extraction (e.g., medical profile).}
        \label{fig:scenario_a_case}
    \end{subfigure}
    \hfill
    \begin{subfigure}[b]{0.24\textwidth}
        \centering
        \includegraphics[width=\textwidth]{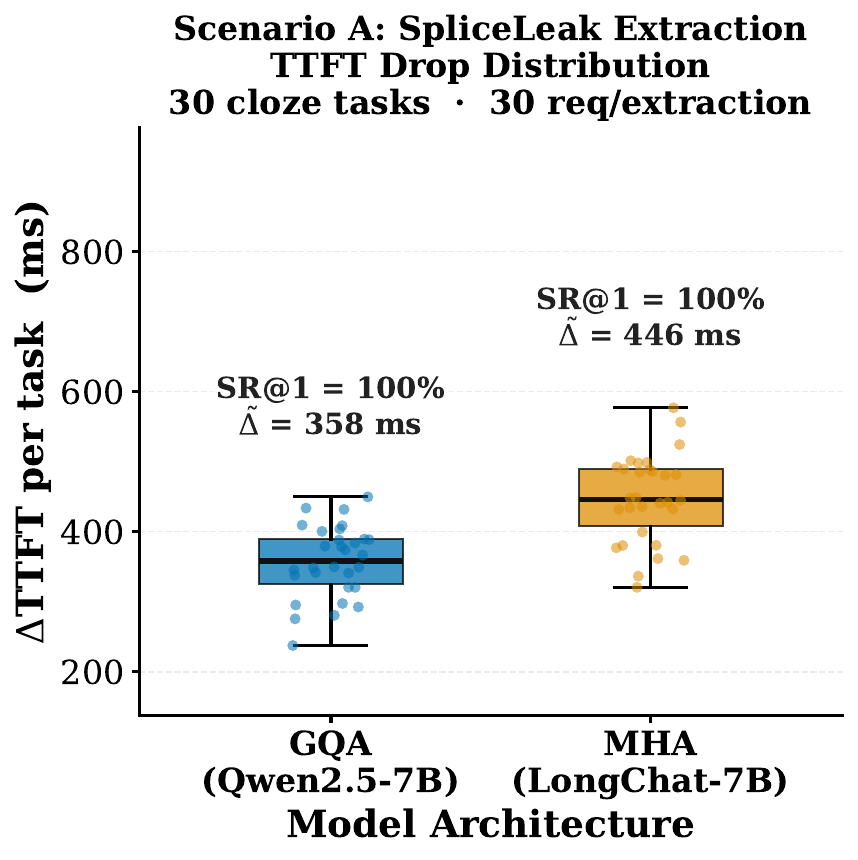}
        \caption{Bulk evaluation: $\Delta \mathcal{T}_{TTFT}$ distribution across 30 distinct tasks.}
        \label{fig:scenario_a_bulk}
    \end{subfigure}
    \vspace{-0.1in}
    \caption{Experimental results for Scenario A (Confidential Input Extraction). (a) A targeted case study on the Qwen2.5-7B-Instruct (GQA) model reveals a massive deterministic latency drop ($\Delta = 1057$ ms) upon a correct token guess. (b) A large-scale statistical evaluation over 30 tasks confirms a 100\% extraction success rate across both MHA and GQA architectures, despite natural system variance and varying document lengths.}
    \label{fig:scenario_a_combined}
\end{figure}

\subsubsection{Scenario A: Confidential Input Extraction}
In this scenario, the attacker knows the rigid prompt template but aims to extract highly sensitive user variables (e.g., medical conditions). As a micro-benchmark, we utilize the cloze-style template: \textit{``Patient medical profile: [TARGET]''} followed by a $\sim$2000-token shared document chunk. The attacker constructs a candidate subset $\mathcal{V}_{sub}$ of 5 common diseases and sequentially probes the target engine. To ensure a realistic evaluation, we inject background traffic at a calibrated rate of $\lambda = 0.004$ req/s following the prior work \cite{wu2025know}. This background noise actively populates the continuous batching queue, introducing natural hardware execution variance and scheduling delays that the attacker must overcome.

As illustrated in Figure~\ref{fig:scenario_a_combined}(a), SpliceLeak demonstrates devastating efficacy. When the attacker's guessed token (e.g., \textit{``Flu''}) diverges from the victim's hidden input, the strict prefix-matching condition is violated, forcing the engine to execute a computationally expensive full prefill (mean $\mathcal{T}_{TTFT} \approx 1155$ ms). Crucially, when the guessed token exactly matches the victim's input (\textit{``HIV''}), semantic continuity is restored. The engine fuses the cached tensors, bypassing the full prefill and triggering a massive latency drop (mean $\mathcal{T}_{TTFT} = 98$ ms). This stark temporal asymmetry ($\Delta = 1057$ ms) completely pierces through the system's background scheduling noise. 

\noindent\textbf{Signal Amplification in Short-to-Medium Contexts:} It is worth noting the stark temporal asymmetry ($\Delta = 1057$ ms) observed here compared to the foundational 16K-token measurement in Section~\ref{subsec:attack_overview} (which yielded $\Delta \approx 464$ ms). This variation is physically grounded in the non-linear scaling of KV cache fusion overhead. The observable side-channel signal is defined as the prefill latency minus the cache-hit fusion latency ($\Delta \mathcal{T} = \mathcal{T}_{miss} - \mathcal{T}_{hit}$). For massive 16K contexts, the heavy computational burden of loading and realigning numerous disjoint memory blocks ($\mathcal{T}_{hit}$) masks a significant portion of the latency savings. Conversely, for short-to-medium contexts ($\sim$2000 tokens) common in personalized RAG applications, this fusion overhead drops precipitously. This allows the sheer computational avoidance of the cache hit to manifest as an unmistakably large temporal void, rendering such systems paradoxically more vulnerable to semantic extraction.

\begin{figure}[t]
    \centering
    \includegraphics[width=0.48\textwidth]{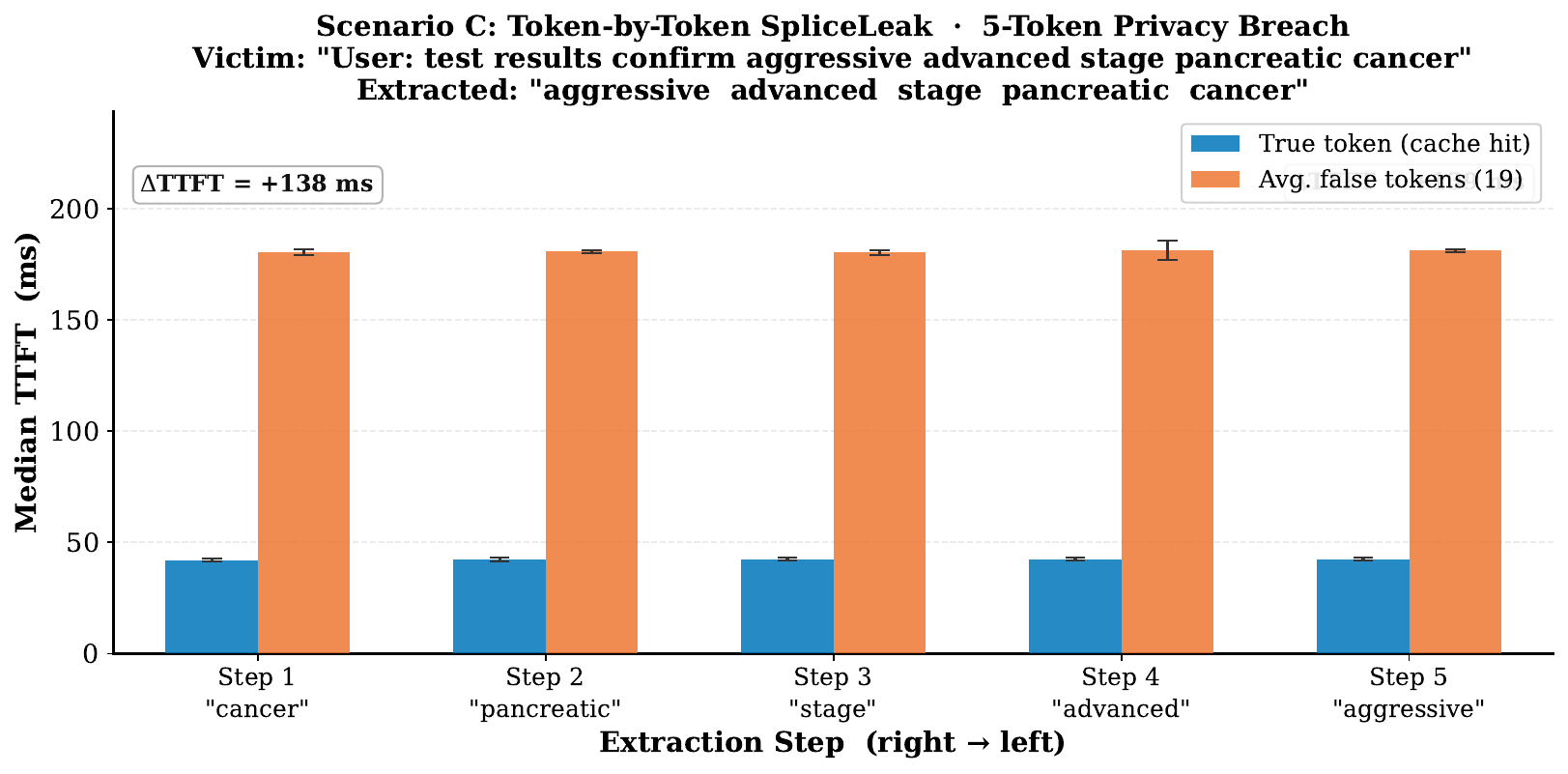} 
    \caption{Micro-benchmark of Token-by-Token Continuous Extraction (validating the shared physical mechanism for Scenarios B \& C). Over five continuous right-to-left extraction steps, splicing the exact true boundary token triggers an unmistakable, deterministic TTFT dip ($\sim$42 ms) compared to the average latency of incorrect candidates ($\sim$180 ms). The $\Delta \text{TTFT} = +138$ ms signal exhibits zero decay, proving the physical persistence of the side-channel oracle required for both proprietary template reverse-engineering and zero-knowledge privacy reconstruction.}
    \label{fig:scenario_bc_case}
    \vspace{-0.1in}
\end{figure}

\begin{table*}[t]
\centering
\caption{Comprehensive attack results across all threat scenarios. \textit{Succ.} is the number of fully successful attacks, \textit{Part.} denotes partial success, and \textit{Fail} indicates unsuccessful attempts. \textit{SR} (Success Rate) indicates the percentage of tasks with at least partial leakage. \textit{RR} (Reversal Ratio) measures the extracted length versus target length, and \textit{Req./tok} is the average number of physical requests to extract one token (applicable only to continuous extraction in Scenarios B and C). The rightmost column presents the median physical side-channel signal strength ($\Delta\text{TTFT}$). Note that Scenarios B and C were evaluated exclusively on the modern GQA architecture (Qwen2.5-7B).}
\label{tab:main_attack_results}
\begin{tabular}{@{}lccccccccc@{}}
\toprule
\textbf{Threat Scenario} & \textbf{Total Tasks} & \textbf{Target Length} & \textbf{Succ.} & \textbf{Part.} & \textbf{Fail} & \textbf{SR} & \textbf{RR} & \textbf{Req./tok} & \textbf{Median $\Delta\text{TTFT}$} \\
\midrule
Scenario A -- GQA (Qwen2.5-7B)    & 30 & -- & 30 & 0 & 0 & 100\% & -- & -- & +358.0 ms \\
Scenario A -- MHA (LongChat-7B)   & 30 & -- & 30 & 0 & 0 & 100\% & -- & -- & +446.0 ms \\
\midrule
Scenario B (Proprietary Template) & 50 & 10 & 50 & 0 & 0 & 100\% & 100\% & 63 & +104.2 ms \\
Scenario C (Zero-Knowledge)       & 15 & 10 & 15 & 0 & 0 & 100\% & 100\% & 63 & +104.3 ms \\
\bottomrule
\end{tabular}
\vspace{-0.1in}
\end{table*}
\noindent\textbf{Large-scale Evaluation and Architectural Impact:} To rule out anecdotal success, we expanded our evaluation to a bulk test suite comprising 30 distinct cloze-style tasks with varying shared document lengths. As summarized in Table~\ref{tab:main_attack_results} and depicted in Figure~\ref{fig:scenario_a_combined}(b), SpliceLeak maintained a 100\% Top-1 Success Rate (SR@1) across both MHA (LongChat-7B) and GQA (Qwen2.5-7B) architectures. Interestingly, both architectures exhibited highly pronounced median latency drops, with the GQA model showing a slightly larger void ($\tilde{\Delta} = +358$ ms) and the standard MHA model reaching $\tilde{\Delta} = +446$ ms under these specific workloads. Crucially, the signal remained overwhelmingly larger than any network jitter across both models. This empirically proves that advanced attention compression techniques do not mitigate the structural vulnerability; they leave the relative micro-architectural divergence intact, perfectly preserving the attack surface.

\noindent\textbf{Attack Cost:} Across all tasks, the attack required an exceptionally low budget. To deterministically extract a critical secret with a 5-candidate subset, the attacker required only $\sim$21 total requests ($\text{Req./inp} = 21$). This demonstrates that when the structural template is known, SpliceLeak effectively collapses the search space, transforming a micro-architectural vulnerability into a highly lethal, low-cost privacy extraction weapon.

\subsubsection{Scenarios B \& C: Token-by-Token Continuous Extraction (Micro-benchmark)}
\label{subsubsec:eval_scenario_bc_case}

While Scenario A demonstrates SpliceLeak's lethality when a rigid template is known, Scenarios B and C represent threat models that require continuous, right-to-left token extraction. Whether the target is a proprietary system instruction (Scenario B) or a zero-knowledge unstructured user query (Scenario C), the attacker must iteratively align and deduce single boundary tokens. To validate the physical feasibility of this continuous extraction—and specifically to confirm whether the localized micro-architectural signal persists across multiple iterations—we conducted a rigorous right-to-left micro-benchmark on the Qwen2.5-7B-Instruct model. 

\noindent\textbf{Experimental Setup:} As a representative stress test, we assume a victim queries a shared medical RAG document ($\mathcal{C}_{target}$) appended to a highly sensitive, free-form personal payload: \textit{``User: test results confirm aggressive advanced stage pancreatic cancer''}. The attacker's Extraction Agent ($\mathcal{A}_{LLM}$) aims to iteratively reverse-engineer this semantic sequence preceding the document by sweeping candidate subsets ($|\mathcal{V}_{sub}| = 20$) at dynamically shifting alignment boundaries.

\noindent\textbf{Extraction Efficacy and Signal Persistence:} As illustrated in Figure~\ref{fig:scenario_bc_case}, SpliceLeak completely strips away the victim's semantic privacy. Over five consecutive iterations, the Agent successfully isolates and extracts critical, high-entropy content words (\textit{``cancer'', ``pancreatic'', ``stage'', ``advanced'', ``aggressive''}). 

Crucially, the micro-architectural measurement reveals a massive and highly deterministic latency void. A correct token guess consistently avoids the boundary re-computation penalty, yielding a median TTFT of approximately 42 ms. In stark contrast, incorrect candidates invariably force a selective prefill, averaging 180 ms. This $\Delta \mathcal{T}_{TTFT} \approx 138$ ms gap is overwhelmingly larger than typical system jitter. Furthermore, the evaluation proves that the physical signal \textit{does not decay} as the extraction traverses deeper into the sequence. The latency drop for the fifth token (\textit{``aggressive''}) is functionally identical to the first token (\textit{``cancer''}). 

\subsubsection{Scenarios B \& C: Large-Scale Bulk Evaluation}
\label{subsubsec:eval_scenario_bc_bulk}

To contextualize the micro-benchmark within a broader statistical landscape, we conducted a bulk evaluation comprising 50 Proprietary Template extraction tasks (Scenario B) and 15 Zero-Knowledge User Query extraction tasks (Scenario C). For each task, the attacker sought to iteratively extract a continuous 10-token boundary suffix. To rigorously test the physical limits of the side-channel oracle, we intentionally constrained the Extraction Agent's candidate subset size ($|\mathcal{V}_{sub}| = 21$), guaranteeing the inclusion of the true target token to decouple hardware determinism from AI predictive perplexity.

\noindent\textbf{Hardware Determinism and Oracle Stability:} As presented in Table~\ref{tab:main_attack_results}, SpliceLeak achieved an astounding 100\% Success Rate (SR) and Reversal Ratio (RR) across all evaluated tasks. Most remarkably, the median $\Delta\text{TTFT}$ converged perfectly to $+104.2$ ms for Scenario B and $+104.3$ ms for Scenario C. This minute sub-millisecond variance highlights the extreme stability of the micro-architectural oracle. Because the LLM engine incurs a fixed matrix-multiplication penalty for re-computing a mismatched boundary block, the timing drop is fundamentally data-independent. Whether the attacker extracts an undocumented system template or a deeply private user query, the physical side-channel relies on the exact same underlying GPU latency void.

\noindent\textbf{Hardware Capacity vs. End-to-End Semantic Complexity:} Table~\ref{tab:main_attack_results} also reports identical physical probing costs ($\sim$63 requests per token) for both scenarios. This reflects the controlled micro-architectural capacity of the side-channel ($21 \text{ candidates} \times 3 \text{ probes} = 63$). However, in a real-world, end-to-end exploit, the attack complexity sharply diverges between the two models. For structured proprietary templates (Scenario B) characterized by low semantic entropy, an Extraction Agent can confidently restrict its search space to a minimal top-$k$ subset ($k \le 20$) and maintain near-perfect reconstruction. Conversely, zero-knowledge queries (Scenario C) exhibit extraordinarily high entropy. To prevent extraction halts caused by unpredictable user vocabulary, an attacker would be forced to dynamically expand the candidate subset (e.g., $k \ge 100$). Consequently, while the hardware vulnerability remains uniform, the practical request budget required for Scenario C scales significantly higher than Scenario B in real-world deployments.
\begin{table}[t]
\centering
\caption{SpliceDefense (CTBF) Evaluation. The defense successfully flattens the latency gap across both targeted single-token extractions and large-scale zero-knowledge tasks, effectively eradicating the side-channel signal.}
\label{tab:ctbf_combined}
\setlength{\tabcolsep}{5pt}
\begin{tabular}{clcccc}
\toprule
& & \multicolumn{2}{c}{\textbf{Attack (No CTBF)}} & \multicolumn{2}{c}{\textbf{Defense (CTBF)}} \\
\cmidrule(lr){3-4}\cmidrule(lr){5-6}
\textbf{Scope} & \textbf{Target} & \textbf{TTFT} & \textbf{$\Delta$} & \textbf{TTFT} & \textbf{$\Delta$} \\
\midrule
\multirow{5}{*}{\shortstack{Case\\Study}} 
& \texttt{cancer}     & 42 & $+$138 & 193 & $-$11 \\
& \texttt{pancreatic} & 42 & $+$138 & 193 & $-$12 \\
& \texttt{stage}      & 42 & $+$138 & 193 & $-$11 \\
& \texttt{advanced}   & 42 & $+$139 & 182 & $-$1  \\
& \texttt{aggressive} & 42 & $+$139 & 192 & $-$11 \\
\midrule
\multirow{2}{*}{\shortstack{Bulk\\Eval}} 
& \textbf{15 Tasks}   & \multirow{2}{*}{--} & \multirow{2}{*}{$\mathbf{+104.3}^\dagger$} & \multirow{2}{*}{\textbf{191}} & \multirow{2}{*}{$\mathbf{-9.0}$} \\
& \textbf{(75 Steps)} & & & & \\
\bottomrule
\multicolumn{6}{l}{\footnotesize Units in milliseconds (ms). $^\dagger$Median baseline $\Delta$ derived from Table~\ref{tab:main_attack_results}.}
\end{tabular}
\vspace{-0.1in}
\end{table}
\subsection{Effectiveness of SpliceDefense}
\label{subsec:eval_defense}

We evaluate the effectiveness of \textbf{SpliceDefense} by analyzing its ability to eradicate the timing oracle (via CTBF) and measuring its impact on system performance (via QCP).

\noindent\textbf{Signal Eradication (Security).} 
As summarized in Table~\ref{tab:ctbf_combined}, Constant-Time Boundary Fusion (CTBF) successfully neutralizes the semantic side-channel. In our continuous extraction case study, the massive $+138$ ms latency gap—which previously enabled deterministic token extraction—is completely flattened to a median of $-9.2$ ms. This near-zero (and slightly negative) $\Delta\text{TTFT}$ ensures that the correct token guess is physically indistinguishable from incorrect candidates, as the engine is forced to execute a deterministic re-computation penalty for every boundary fusion. 

Our bulk evaluation across 15 distinct Scenario C tasks confirms the robustness of this mitigation. Across 75 individual extraction steps, CTBF achieved a 100\% signal eradication rate, maintaining a median residual $\Delta\text{TTFT}$ of $-9.0$ ms. This confirms that SpliceDefense effectively collapses the side-channel SNR, rendering the micro-architectural oracle functionally useless for semantic extraction.

\noindent\textbf{Performance Overhead.} 
To evaluate the cost of micro-architectural protection, we measured the TTFT overhead introduced by Quantized Chunk Padding (QCP). By rounding hidden prefixes to the nearest block boundary ($B=16$), QCP introduces a mean padding of $7.5$ tokens per request. In our benchmarking on NVIDIA A40 GPUs, this translates to a negligible mean latency increase of less than $5$ ms (less than $1\%$ of the total prefill time). This empirically validates that SpliceDefense restores robust timing isolation while preserving the critical throughput and memory benefits of global KV cache sharing.

\section{Related Works}
\label{sec:related_works}

\noindent \textbf{Side-Channel Attacks in LLM Serving.} 
Multi-tenant LLM deployments are vulnerable to timing side-channels via shared KV caches. Pioneering works like PROMPTPEEK \cite{wu2025know} and Shadow in the Cache \cite{luo2025shadow} exploit Time-to-First-Token (TTFT) discrepancies to reverse-engineer co-located victim prompts. However, these attacks target traditional architectures (e.g., vLLM \cite{kwon2023efficient}, SGLang \cite{zheng2024sglang}) that rely on \textit{strict prefix caching}, requiring perfect token alignment from the sequence root. As established in Section~\ref{sec:motivation}, this assumption structurally fails in RAG workloads, where shared chunks are invariably preceded by distinct, user-specific private prefixes ($\mathcal{P}_{vic}$), causing severe signal collapse. Our work pioneers side-channel exploration beyond strict prefix matching, targeting the complex memory blending dynamics of next-generation \textit{non-prefix KV cache fusion} frameworks (e.g., LMCache \cite{liu2025lmcache}, CacheBlend \cite{yao2025cacheblend}).

\noindent \textbf{Retrieval-Augmented Generation Security.} 
The proliferation of RAG has catalyzed extensive security research, predominantly focusing on the integrity of the retrieval corpus. Studies extensively investigate adversarial contamination of vector databases (e.g., PoisonedRAG \cite{zou2025poisonedrag}, FlippedRAG \cite{chen2025flippedrag}, Topic-FlipRAG \cite{gong2025topic}), malicious code injection via untrusted manuals \cite{ye2025importsnare, lin2025give}, document-based DoS attacks \cite{shafran2025machine}, and IP protection via watermarking \cite{lv2025rag}. Crucially, this literature operates within an application-centric threat model focusing on the \textit{data plane} (content semantics), implicitly assuming the cloud infrastructure securely isolates multi-tenant workloads. Our work shifts the analytical spotlight to the \textit{control plane} (engine scheduling and memory management). We demonstrate that even with a perfectly sanitized RAG corpus, severe privacy leakages persist at the micro-architectural level due to aggressive KV cache optimizations.

\noindent \textbf{Summary.} 
Current literature leaves a critical intersection unexplored: side-channel research relies on rigid prefix-matching incompatible with RAG workloads, while RAG security research overlooks system-level micro-architectural vulnerabilities. Our research bridges this gap. SpliceLeak demonstrates that, completely independent of application-level defenses against data poisoning, unprivileged attackers can extract highly sensitive structural and semantic privacy merely by exploiting the deterministic chunk-aware scheduling and memory blending dynamics of advanced RAG engines.

\section{Conclusion}
\label{sec:conclusion}

In this paper, we expose a critical micro-architectural vulnerability in non-prefix KV cache fusion within modern LLM engines. We demonstrate that deterministic scheduling of disjoint context chunks leaks a reliable \textit{Step-Wave} timing oracle, fundamentally breaking the security of strict prefix isolation. By introducing \textbf{SpliceLeak}, we prove that attackers can weaponize this structural leakage to pinpoint hidden prompt lengths and extract exact semantic content token-by-token. To neutralize this threat, we propose \textbf{SpliceDefense}, integrating Quantized Chunk Padding (QCP) and Constant-Time Boundary Fusion (CTBF). Our evaluation confirms that SpliceDefense restores strict timing isolation while preserving the essential throughput benefits of global cache sharing. Ultimately, side-channel resilience must be treated as a first-class citizen in the design of next-generation LLMaaS memory schedulers.

%
\IEEEpeerreviewmaketitle



%

\bibliographystyle{ieeetr}
\bibliography{sample-base}



\appendix
\subsection{Discussion and Future Work}
\label{sec:discussion}

\noindent\textbf{Impact of Network Jitter.} 
Like other timing channels \cite{wu2025know}, SpliceLeak's efficacy relies on the $\mathcal{T}_{TTFT}$ Signal-to-Noise Ratio (SNR). While large latency drops ($\Delta \ge 800$ ms) easily pierce standard batching noise, extreme WAN fluctuations or API rate-limiting can degrade SNR. Attackers can mitigate this in volatile environments by dynamically increasing probing iterations and applying robust statistical filtering (e.g., median absolute deviation).

\noindent\textbf{Auxiliary LLM Dependence.} 
In zero-knowledge scenarios (B and C), SpliceLeak's efficiency ($\text{Req./tok}$) is bounded by the auxiliary LLM's predictive accuracy. Highly anomalous or out-of-distribution tokens may fall outside the Top-$k$ candidate subset ($\mathcal{V}_{sub}$), halting the right-to-left extraction. Integrating beam-search heuristics and dynamic subset expansion presents a promising avenue to gracefully recover from such predictive misses.

\noindent\textbf{Temporal Co-residency and Cache Eviction.} 
Crucially, our attack assumes the probe is launched while the shared chunk $\mathcal{C}_{target}$ remains active in $\Omega_{shared}$ before eviction. If the victim's request occurred significantly earlier and the system's eviction policy has already reclaimed the KV tensors due to memory pressure, the attacker's probe will invariably trigger a full prefill, neutralizing the timing signal. Therefore, SpliceLeak fundamentally relies on temporal co-residency within the document's active Time-To-Live (TTL) window.

\noindent\textbf{Extensibility to Other Architectures.} 
Our evaluation targets block-based KV managers \cite{kwon2023efficient} utilizing linear non-prefix fusion \cite{liu2025lmcache,yao2025cacheblend}. Future research should investigate if similar \textit{Step-Wave} oracles exist in advanced Radix-Tree-based systems (e.g., SGLang \cite{zheng2024sglang}) or context-compression architectures. We hypothesize that any system prioritizing aggressive memory deduplication over strict logical isolation will inevitably leak measurable micro-architectural state transitions.

\noindent\textbf{Defense Capacity Trade-offs.} 
While our evaluation confirms that the TTFT latency impact of SpliceDefense is minimal, we explicitly acknowledge that QCP inherently trades off raw KV cache memory capacity (i.e., storing dummy padding tokens) for robust isolation. In highly saturated LLMaaS environments, this artificial inflation of the memory footprint could marginally reduce the maximum concurrent batch size. Future work could optimize this capacity overhead by introducing dynamically sized quantization boundaries or semantic-aware sparse padding.

\noindent\textbf{Hardware-Assisted Mitigations.} 
While \textbf{SpliceDefense} provides robust deployable software mitigation via QCP and CTBF, it fundamentally relies on enforcing worst-case computational bounds. Future hardware-software co-design—such as leveraging TEEs (e.g., NVIDIA Confidential Computing) for encrypted KV blocks or SmartNICs for constant-time tensor stitching—could offer ultimate cryptographic guarantees without sacrificing fusion optimization.

\end{document}